\documentclass[aip,graphicx]{revtex4-1}%word document

\draft % marks overfull lines with a black rule on the right

\usepackage{graphicx}% Include figure files
\usepackage{dcolumn}% Align table columns on decimal point
\usepackage{bm}% bold math
\usepackage[mathlines]{lineno}% Enable numbering of text and display math
\usepackage{amssymb}

\usepackage[utf8]{inputenc}
\usepackage[T1]{fontenc}
\usepackage{mathptmx}
\usepackage{etoolbox}
\usepackage{xcolor}

\makeatletter
\def\@email#1#2{%
 \endgroup
 \patchcmd{\titleblock@produce}
  {\frontmatter@RRAPformat}  {\frontmatter@RRAPformat{\produce@RRAP{*#1\href{mailto:#2}{#2}}}\frontmatter@RRAPformat}
  {}{}
}%
\makeatother
\begin{document}

\preprint{}% Use the \preprint command to place your local institutional report number on the title page in preprint mode.
\graphicspath{{./images/}}

\title[A practical guide to light-sheet microscopy for nanoscale imaging]{A practical guide to light-sheet microscopy for nanoscale imaging:\\Looking beyond the cell}

\author{Stephanie N. Kramer}
\author{Jeanpun Antarasen}
\author{Cole R. Reinholt}
\affiliation{Department of Physics, Case Western Reserve University, Rockefeller Building, 2076 Adelbert Road, Cleveland, Ohio 44106, USA}
\author{Lydia Kisley}%
\affiliation{Department of Physics, Case Western Reserve University, Rockefeller Building, 2076 Adelbert Road, Cleveland, Ohio 44106, USA}
\affiliation{Department of Chemistry, Case Western Reserve University, Clapp Hall, 2080 Adelbert Road, Cleveland, Ohio 44106, USA}
\email{lydia.kisley@case.edu}

\date{\today}

\begin{abstract}
We present a comprehensive guide to light-sheet microscopy (LSM) to assist scientists in navigating the practical implementation of this microscopy technique. Emphasizing the applicability of LSM to image both static microscale and nanoscale features, as well as diffusion dynamics, we present the fundamental concepts of microscopy, progressing through beam profile considerations, to image reconstruction. We outline key practical decisions in constructing a home-built system and provide insight into the alignment and calibration processes. We briefly discuss the conditions necessary for constructing a continuous 3D image and introduce our home-built code for data analysis. By providing this guide, we aim to alleviate the challenges associated with designing and constructing LSM systems and offer scientists new to LSM a valuable resource in navigating this complex field.
\end{abstract}


\maketitle

\section{\label{sec:Intro}Introduction}
Fluorescence microscopy techniques offer non-invasive means to study biological samples \textit{in situ}. An ideal fluorescence microscope should possess high sensitivity, signal-to-background ratio (SBR), and spatial resolution, along with fast acquisition speeds and minimal photodamage. A sensitive microscope typically exhibits lower background intensity relative to the sample emission, allowing for detection of weaker signals amidst background noise. Fluorescence microscopy inherently benefits from high contrast imaging due to the substantial quantum yield of fluorophores, enhancing image quality and enabling precise visualization of low-abundance molecules or events.\cite{verveer_high-resolution_2007,stelzer_light-sheet_2015,olarte_light-sheet_2018} High spatial resolution is crucial for resolving fine structural details, particularly at the subcellular level. Fast image acquisition with low source intensity, facilitated using conventional detectors like fast frame rate cameras, can help mitigate photodamage. Being able to collect at higher frame rates also improves the resolvable timescales for more dynamics-based analysis.
\begin{figure}
\includegraphics[width=0.5\textwidth]{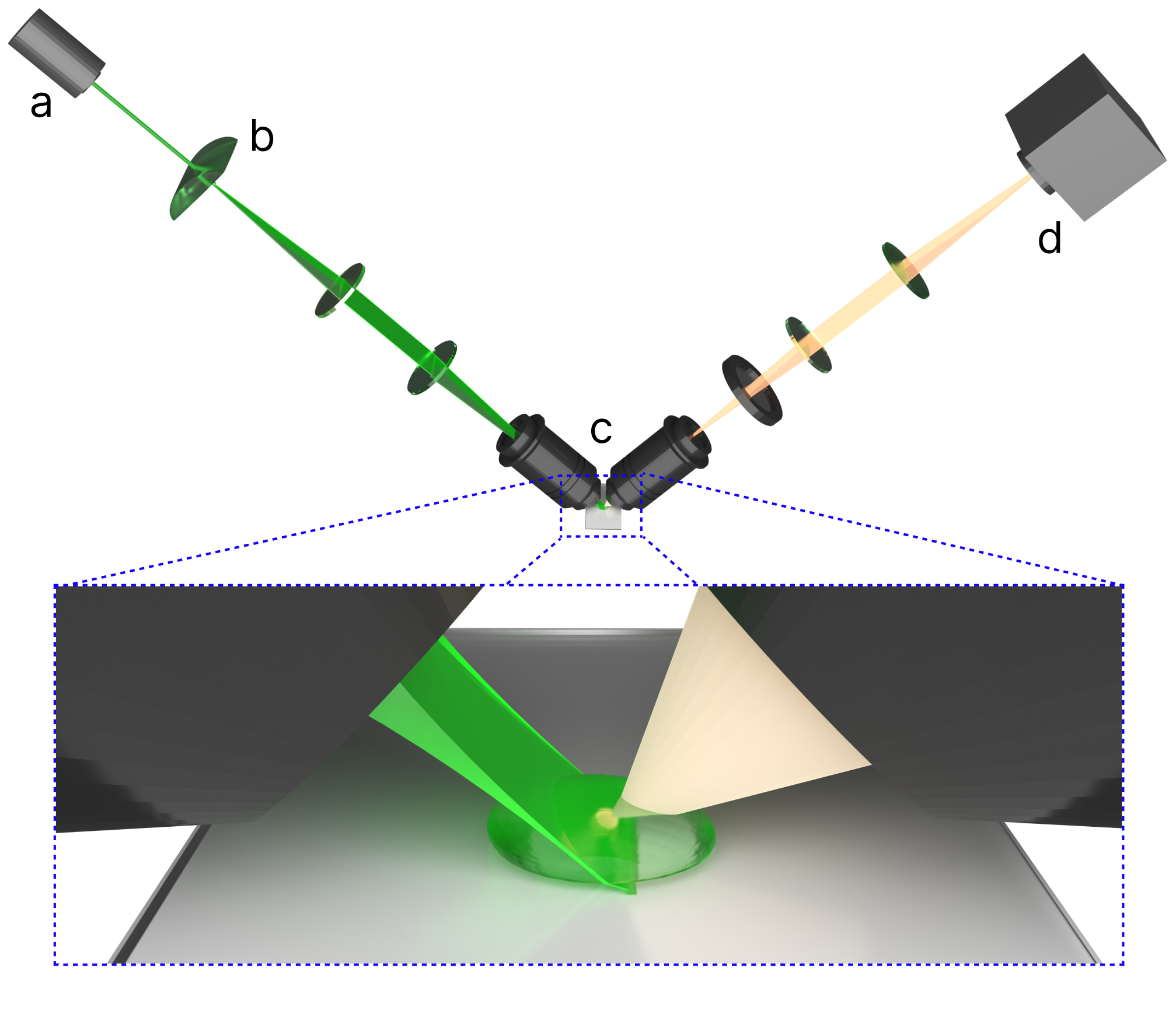}
\caption{\label{fig:basicLSM} Basic diagram of a light-sheet microscope set-up. The excitation source of a LSM set-up is generally a laser, the beam profile of which is either Gaussian, Bessel, or Airy in nature (a). Light-sheet formation can be accomplished using cylindrical lenses or via scanning mirrors (b). Most LSMs use a two-objective system with separate illumination and detection objectives (c). The illumination objective focuses a thin sheet through a sample and the detection objective collects the sample emission. The detectors for LSMs are usually either electron multiplying charge-coupled device (EMCCD) or scientific complementary metal-oxide-semiconductor (sCMOS) cameras (d). Inset: The thin sheet allows for lower background signal as well as less photodegradation of the sample. Sample mounting will be determined based on the objective system chosen as well as the overall geometry of the LSM.}
\end{figure}

Epifluorescence\cite{sieracki_detection_1985,bai_full-color_2020,pasulka_visualization_2021} and scanning confocal microscopy\cite{shotton_confocal_1989,bayguinov_modern_2018,elliott_confocal_2020} are common fluorescence methods used for biological imaging, yet they have limitations in optical sectioning and timescale resolution, respectively. Epifluorescence microscopy illuminates the entire sample, resulting in lower contrast caused by the out-of-focus signal. The additional excited fluorophores outside of the focal plane also limits the optical sectioning and increases photodamage.\cite{li_fluorescence_2004,thorn_quick_2016} However, epifluorescence microscopy can utilize the fast frame rates of the chosen detector to increase the resolvable timescale. Scanning confocal microscopy offers improved optical sectioning due to the incorporation of a pinhole  that blocks light outside of the focal volume, and  also improves the image contrast.\cite{li_fluorescence_2004,thorn_quick_2016} The single-point measurement still induces notable photodamage due to requiring multiple scanning iterations to construct an image. However, the smaller sections do limit the overall photodamage compared to widefield epifluorescence method. Considering the constraints of both epifluorescence and scanning confocal microscopy, including limited 3D sectioning capabilities, potential for inducing sample photodamage, and susceptibility to detecting excess signal beyond the focal plane, it becomes imperative to seek out and adopt an enhanced imaging technique for more precise and comprehensive analysis.

Light-sheet microscopy (LSM) has gained acclamation in biological imaging due to the selectively of the illuminating focal plane, reducing photodamage and improving SBR (Fig.~\ref{fig:basicLSM}).\cite{huisken_optical_2004, stelzer_light-sheet_2015,olarte_light-sheet_2018,reynaud_light_2008}
The selective illumination is due in part to the thin focal volume of the sheet with beam thicknesses reaching as small as 400 nm\cite{poola_light_2019,malivert_active_2022,capoulade_quantitative_2011} in specialized set-ups such as lattice LSM while more conventional set-ups afford thicknesses of a few $\mu$m.\cite{takanezawa_wide_2021,sapoznik_versatile_2020} The ability of LSM to capture entire planes simultaneously enhances image acquisition speed, making it suitable for dynamic imaging experiments over extended periods such as monitoring zebrafish embryo development,\cite{trivedi_dynamic_2015,bernardello_light-sheet_2021} cell migration,\cite{ritter_actin_2015,schmid_high-speed_2013} or neuronal activity\cite{ahrens_whole-brain_2013,hosny_planar_2020} where the timescales range from ms to hours. In addition to these dynamic samples, LSM has also been used to characterize structural features at high resolution within samples ranging from zebrafish\cite{yang_dual-slit_2015,fei_cardiac_2016} to single cells.\cite{welf_quantitative_2016,wu_light_2013} While this is an impressive display of imaging, all listed example analytes have features on the order of $\mu$m. Therefore, a 3D reconstruction of most biological samples can be acquired using commercially available LSM set-ups.

LSMs with super-resolution capabilities are necessary for imaging nanoscale features at or below the diffraction limit (< 250 nm) or dynamic samples at timescales faster than a few ms. Generally, to achieve super-resolution, a more specialized system is needed which will require a custom set-up to be constructed. Designing such systems requires careful consideration of optical components and alignment processes. In our guide, we detail considerations for building a home-built LSM for imaging non-traditional systems at nanoscales. We discuss optical component selection, alignment procedures, and our design choices for measuring biomolecule diffusion dynamics within an extracellular matrix (ECM) analogue, covering alignment, calibration, data acquisition, and 3D reconstruction processes.

\section{\label{sec:Config}Microscope design components to consider prior to construction} 
Each light-sheet set-up is unique, tailored to both the chosen sample and the desired spatiotemporal resolutions. When beginning to consider the configuration of the microscope, first consider the characteristics of the sample, as they influence subsequent component choices. For instance, if the sample requires a specific orientation, this influences objective selection and the design of the sample chamber. For example, imaging zebrafish necessitates adherence to a substrate to minimize movement while for smaller entities like cells, immersion in a suitable buffer solution and substrate adherence are essential. Dynamic data collection, such as protein diffusion dynamics, imposes distinct experimental requirements, notably in resolving diffusion timescales (10-100 $\mu$m$^2/s$), compared to static samples. Regardless of the sample, identifying necessary experimental conditions prior to construction is crucial. In this section, we outline hardware considerations encompassing laser properties, sheet dimensions, objective geometry, sample immersion and orientation, and camera sensitivity and timescales.

\subsection{\label{sec:Beams}Beam profiles that have been used in LSM}
\begin{figure}
\includegraphics[width=0.5\textwidth]{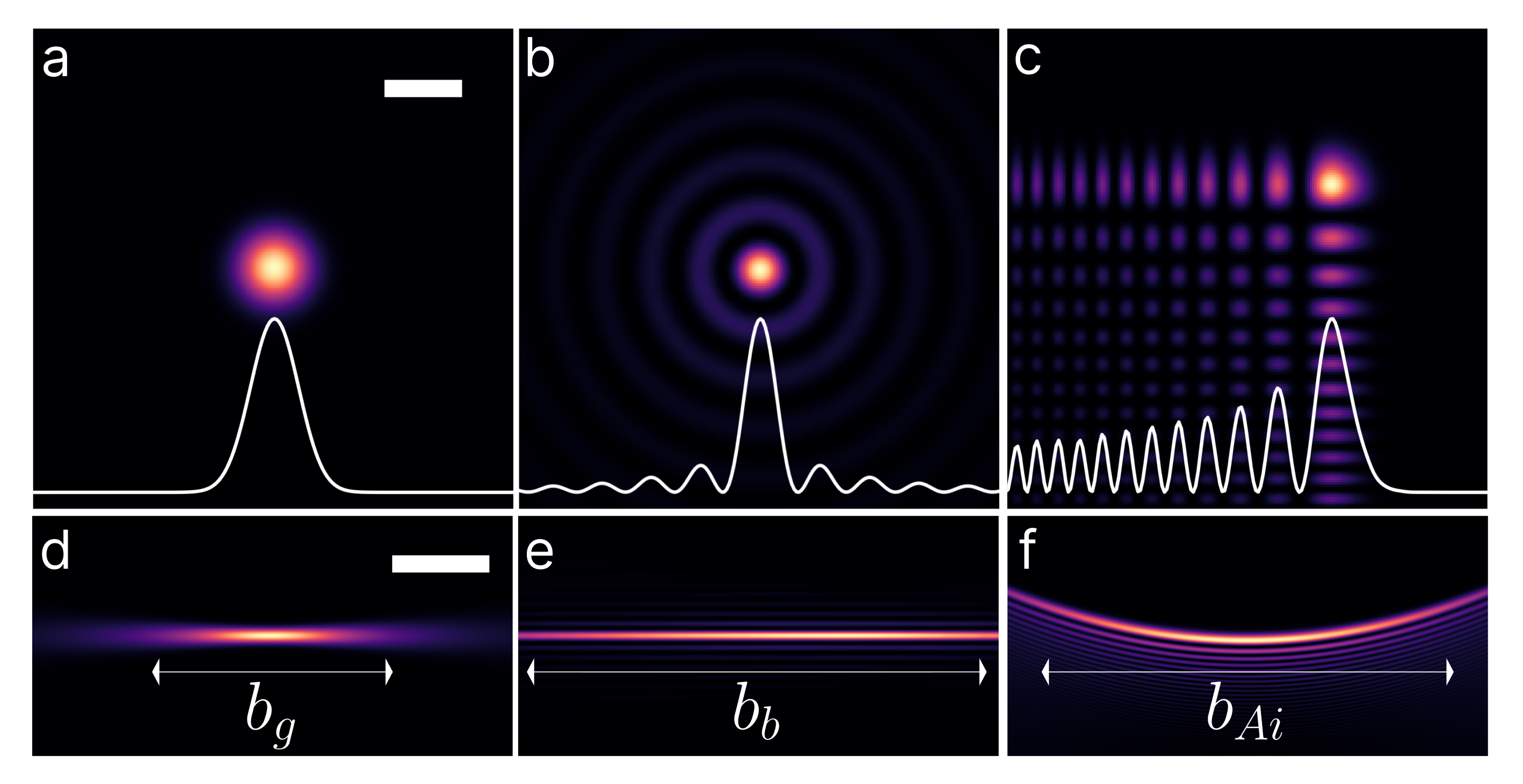}
\caption{\label{fig:beamtypes}Profiles of Gaussian (a,d), Bessel (b,e) and Airy (c,f) beams. Here we show the x-y cross-sectional profiles at $z=0$ (a-c, intensity profiles inset) and the axial profiles (d-f). The top row illustrates the overall difference in illumination between each beam, while the bottom row shows the differences in the confocal parameters. A Gaussian beam is generated from the theoretical model using Eq.~(\ref{eq:GBint}). Bessel beams are generated through a Gaussian beam propagating through an axicon, displaying energy dissipated along the lateral axis. The Airy beam is also simulated using the theoretical model based on the propagation of a Gaussian beam through a cubic phase modulator. We use the Lightpipes module in Python to demonstrate the optical phenomena of each beam.\cite{van_goor_lightpipes_2017} The scale bar for (a-c) is 25 mm, while for (d-f), it is 25 cm.}
\end{figure}

Three beam profiles— Gaussian,\cite{capoulade_quantitative_2011,ahrens_whole-brain_2013} Bessel,\cite{planchon_rapid_2011,takanezawa_wide_2021} and Airy\cite{vettenburg_light-sheet_2014,corsetti_widefield_2020}—have proven effective in LSM (Fig.~\ref{fig:beamtypes}). Each offers distinct advantages, although no single profile suits every experiment perfectly. The primary considerations for profile selection are the desired axial resolution and lateral field of view (FOV). Here, we will explore the characteristics, advantages, and drawbacks of these profiles for LSM set-ups. Notably, we will omit discussion of less conventional options like lattice profiles, which entail complexities in integration into specialized LSM set-ups, requiring custom objectives and programmed coordination between electronically controlled mirrors, objectives, and the sample stage.\cite{chen_lattice_2014,malivert_active_2022,shi_smart_2024,moore_multi-functional_2021} Commercial lattice LSM systems are available to address such needs.\cite{3i_lattice_nodate,zeiss_elyra_nodate,zeiss_lattice_nodate} Among the three discussed profiles, Gaussian profiles are the simplest, with Bessel and Airy profiles typically formed from Gaussian beams. While the intensity profile of a Gaussian beam can be solved analytically in all three dimensions, such analysis is not straightforward for Bessel and Airy profiles despite their derivation from Gaussian beams. Thus, we simplify the analysis by considering two dimensions for Bessel and Airy intensity profiles.

\subsubsection{\label{sec:Gaussian}Gaussian}
Gaussian beams are common due to their simplicity and the prevalence of lasers with Gaussian or closely approximated two-dimensional intensity profiles (Fig.~\ref{fig:beamtypes}a). The beam profile is described by a Gaussian function
\begin{eqnarray}
I(r,z) = I_{0} \frac{2}{\pi \omega_{0g}^2(1+\frac{z}{z_{Rg}})} \exp(\frac{-2r^2}{\omega_{0g}^2(1+\frac{z}{z_{Rg}})}),
\label{eq:GBint}
\end{eqnarray}
where $I_{0}$ is the maximum intensity of the beam at $z = 0$, $\omega_{0g}$ the radius of the beam where $I = I_{0}/e^2$ (13.5\%), $z_{Rg}$ the Rayleigh range, and $r$ the radial distance from the center axis of the beam.\cite{glaser_multidirectional_2018} The beam waist $\omega_{0g}$ can be characterized using Eq.~(\ref{eq:GBss})
\begin{eqnarray}
\omega_{0g} = \frac{\lambda}{\pi\theta},
\label{eq:GBss}
\end{eqnarray}
where $\lambda$ is the laser wavelength and $\theta$ the divergence angle.

When opting for a Gaussian beam, achieving both reasonable axial resolution, determined by $\omega_{0g}$, and a wide FOV, based on propagation length, involves a trade-off. This balance is described by the Rayleigh range, defined as the distance from $\omega_{0g}$ along the illumination axis where $\omega_{z}\leq\sqrt{2}$$\omega_{0g}$. The Rayleigh range is calculated via Eq.~(\ref{eq:GBrr}) with refractive index $n$, but $z_{Rg}$ represents only half of the full propagation length. Therefore, doubling $z_{Rg}$ yields the full propagation length (Fig.~\ref{fig:beamtypes}d), known as the confocal parameter $b_{g}$ (Eq.~(\ref{eq:GBdof}))
\begin{eqnarray}
z_{Rg} = \frac{\pi\omega_{0g}^2 n}{\lambda},
\label{eq:GBrr}
\end{eqnarray}
\begin{eqnarray}
b_{g} = 2z_{Rg}.
\label{eq:GBdof}
\end{eqnarray}

Gaussian beams with narrow waists offer higher axial resolution but have smaller propagation lengths along the illumination axis. While the sectioning will be better in these systems, it will inevitably limit the possible FOV when imaging. Conversely, wider beam waists sacrifice axial resolution for increased propagation lengths and FOVs.\cite{takanezawa_wide_2021,corsetti_light_2019}
Additionally, Gaussian beams encounter scattering and absorption in biological samples, resulting in irregular patterns like shadows and streaks as light propagates through tissues.\cite{rohrbach_artifacts_2009,royer_practical_2018,huisken_even_2007} The penetration depth inside tissues is thus decreased, making imaging the entire volume more challenging for thicker samples.

\subsubsection{\label{sec:Bessel}Bessel}
A Bessel beam, unlike a Gaussian beam, exhibits non-diffracting properties, maintaining its waist size without spreading out as it propagates.\cite{mcgloin_bessel_2005,khonina_bessel_2020,tumkur_nondiffractive_2021} To use this beam profile, it must be generated from a Gaussian laser beam, a process achievable through various methods such as employing a diffraction grating\cite{jimenez_acoustic_2014} or using an axicon.\cite{takanezawa_wide_2021} An axicon is an optic that produces a line focus along the optical axis rather than a point focus.\cite{sochacki_nonparaxial_1992,herman_production_1991} Various optical elements, such as a narrow annular aperture,\cite{durnin_diffraction-free_1987} refractive cone,\cite{mcleod_axicon_1954} doublet lens,\cite{burvall_simple_2004} conical mirror,\cite{fujiwara_optical_1962} or circular grating\cite{perez_diffraction_1986} can serve as an axion. Regardless of the method used for beam formation, the intensity profile of an ideal Bessel beam is described by the function
\begin{eqnarray}
I(r,z) \propto J_{0}^2(k_{r}r),
\label{eq:BBint}
\end{eqnarray}
where $E$ is the electric field, $J_{0}$ a Bessel function of zeroth order, $k_{r}$ the radial wave vector, and $r$, $\phi$, and $z$ the radial, polar, and longitudinal coordinates, respectively. However, in practical applications, it is not possible to generate infinitely intense non-diffractive beams. This tutorial demonstrates the generation of a Bessel beam by propagating a Gaussian beam through an axicon, resulting in a finite lateral intensity profile.

Bessel beams produced from an axicon exhibit multiple concentric rings around the core (Fig.~\ref{fig:beamtypes}b), unlike an ideal Gaussian beam with a single mode. These concentric rings maintain consistent intensity within a certain range $b_{b}$, analogous to the confocal parameter albeit analytical derived differently
\begin{eqnarray}
b_{b} = \frac{\omega_{og}}{(n-1)\alpha}.
\label{eq:BBrr}
\end{eqnarray}
Here, $b_{b}$ is expressed as a function related to the axicon top angle $\alpha$ and refractive index $n$.\cite{khonina_bessel_2020} In the case of a Bessel beam generated from a circular aperture, the Rayleigh range and beam waist can be analytically calculated.\cite{tumkur_nondiffractive_2021}

In addition to maintaining axial resolution while propagating over longer distances (Fig.~\ref{fig:beamtypes}e), Bessel beams penetrate scattering media more deeply compared to Gaussian beams. Bessel beams possess the ability to "self-heal" in scattering tissue, allowing for increased penetration depth and retention of beam shape.\cite{fahrbach_microscopy_2010,khonina_bessel_2020,corsetti_light_2019,takanezawa_wide_2021} This difference in penetration depth was demonstrated by Purnapatra \textit{et al}. by imaging fluorescent polymer-coated yeast cells suspended in a tissue-like gel.\cite{purnapatra_spatial_2012} Here, the Bessel beam penetrated to approximately 650 $\mu$m compared to the limit of the Gaussian beam of around 200 $\mu$m. However, it is worth noting that the overall image contrast collected with a Bessel beam may be reduced compared to a Gaussian beam due to the multi-lobe nature of a Bessel beam, potentially resulting in out-of-focus excitation.\cite{deng_enhancement_2022,remacha_how_2020}

\subsubsection{\label{sec:Airy}Airy}
An Airy beam is non-diffracting and typically generated from a Gaussian profile\cite{morris_propagation_2009}, similar to a Bessel beam.  Conventionally, Airy beams are formed using a spatial light modulator (SLM), a device capable of applying a spatially varying phase pattern to light, enabling a more compact Airy-based LSM set-up compared to lens-based methods.\cite{latychevskaia_creating_2016,ren_non-diffracting_2021} However, Airy beams can still be formed using lenses.\cite{papazoglou_tunable_2010,hosny_planar_2020} Unlike symmetric Gaussian or Bessel profiles, the intensity profile of an Airy beam is parabolic (Fig.~\ref{fig:beamtypes}f) and asymmetric in nature (Fig.~\ref{fig:beamtypes}c). While an ideal Airy beam possesses infinite power, experimentally it exhibits finite energy and is best described by an exponentially apertured ideal Airy beam.\cite{morris_propagation_2009} It is then possible to numerically propagate the beam in the z-direction. To create an Airy beam from a Gaussian beam, a cubic phase generated by the SLM is utilized. The intensity profile of the Airy beam can be theoretically expressed as
\begin{eqnarray}
I(x, y, z) = \exp \left(2a(\frac{x}{x_0} - \frac{z^2}{2k^2x_0^4})\right) \exp \left(2a(\frac{y}{x_0}-\frac{z^2}{2k^2x_0^4}) \right) \nonumber\\
\left( Ai(\frac{x}{x_0}-\frac{z^2}{4k^2x_0^4}+i\frac{az}{kx_0^2}) Ai(\frac{y}{x_0}-\frac{z^2}{4k^2x_0^4}+i\frac{az}{kx_0^2}) \right)^2, \nonumber\\
\label{eq:ABint}
\end{eqnarray}
where $Ai(x)$ is the Airy function ($\frac{d^2 x}{dx^2}-x\cdot y = 0$), $x_{0}$ the characteristic length, $a$ the apodization factor which describes the amplitude decrease relative to the radial coordinate of the pupil, and $k = \frac{2\pi}{\lambda}$ the wave number.\cite{taege_design_2022}

In Airy LSM, the propagation length can be derived based on the full-width half maximum (FWHM) of the maximum intensity of the main lobe of the beam, akin to the definition of $z_G$ and $z_B$
\begin{eqnarray}
b_{Ai} = 6\sqrt{2 \ln 2}\frac{\lambda}{n\gamma}.
\end{eqnarray}
The parameter $\gamma$, which determines $a$ and $x_0$, can be optimized based on the numerical aperture of the system or as the ratio of the focal length to the waist of the Gaussian beam $w_{0g}$ for a specific $b_{Ai}$.\cite{taege_design_2022}

Similar to Bessel beams, Airy beams maintain comparable axial resolution and exhibit "self-healing," enabling increased maximum depth of penetration compared to Gaussian beams.\cite{mazilu_light_2010,broky_self-healing_2008,corsetti_light_2019,taege_design_2022} Nylk \textit{et al}. observed that Airy beams could penetrate up to approximately 330 $\mu$m into mouse brain tissue, whereas Gaussian beams reached only about 250 $\mu$m.\cite{nylk_enhancement_2016} However, the depth achieved with Airy beams, although significant, is still less than the approximate 650 $\mu$m penetration depth reported for Bessel beams by Purnapatra \textit{et al}..\cite{purnapatra_spatial_2012,nylk_enhancement_2016} Additionally, Airy beams may yield lower contrast images due to their multi-lobe intensity structure as well as necessitate the use of deconvolution methods to correct for asymmetric side-lobe structures.\cite{deng_enhancement_2022, remacha_how_2020,vettenburg_light-sheet_2014}

\subsection{\label{sec:Sheet}Sheet dimensions determine the microscope sectioning capabilities}
In this section, we take a close look at the important considerations when gathering components for constructing a microscope, ranging from optical parts to the practicalities of assembly. Additionally, we describe some calculations to help estimate the dimensions of the light-sheet based on the planned optical path. These calculations are crucial for optimizing a LSM set-up and ensuring it meets the desired imaging needs.

\subsubsection{\label{sec:Method}Static or dynamic sheets can be formed}
Light-sheets are created through the manipulation of Gaussian, Bessel, or Airy beams.\cite{olarte_light-sheet_2018,hosny_planar_2020,luna-palacios_multicolor_2022} Cylindrical lenses focus the excitation beam into a light-sheet in selective-plane illumination microscopy (SPIM),\cite{huisken_optical_2004,huisken_even_2007} while in digitally scanned laser light-sheet microscopy (DSLM), a virtual light-sheet is formed by rapidly moving a focused beam at the focal plane of the detection lens.\cite{keller_reconstruction_2008,keller_digital_2010}

\begin{figure}
\includegraphics[width=0.5\textwidth]{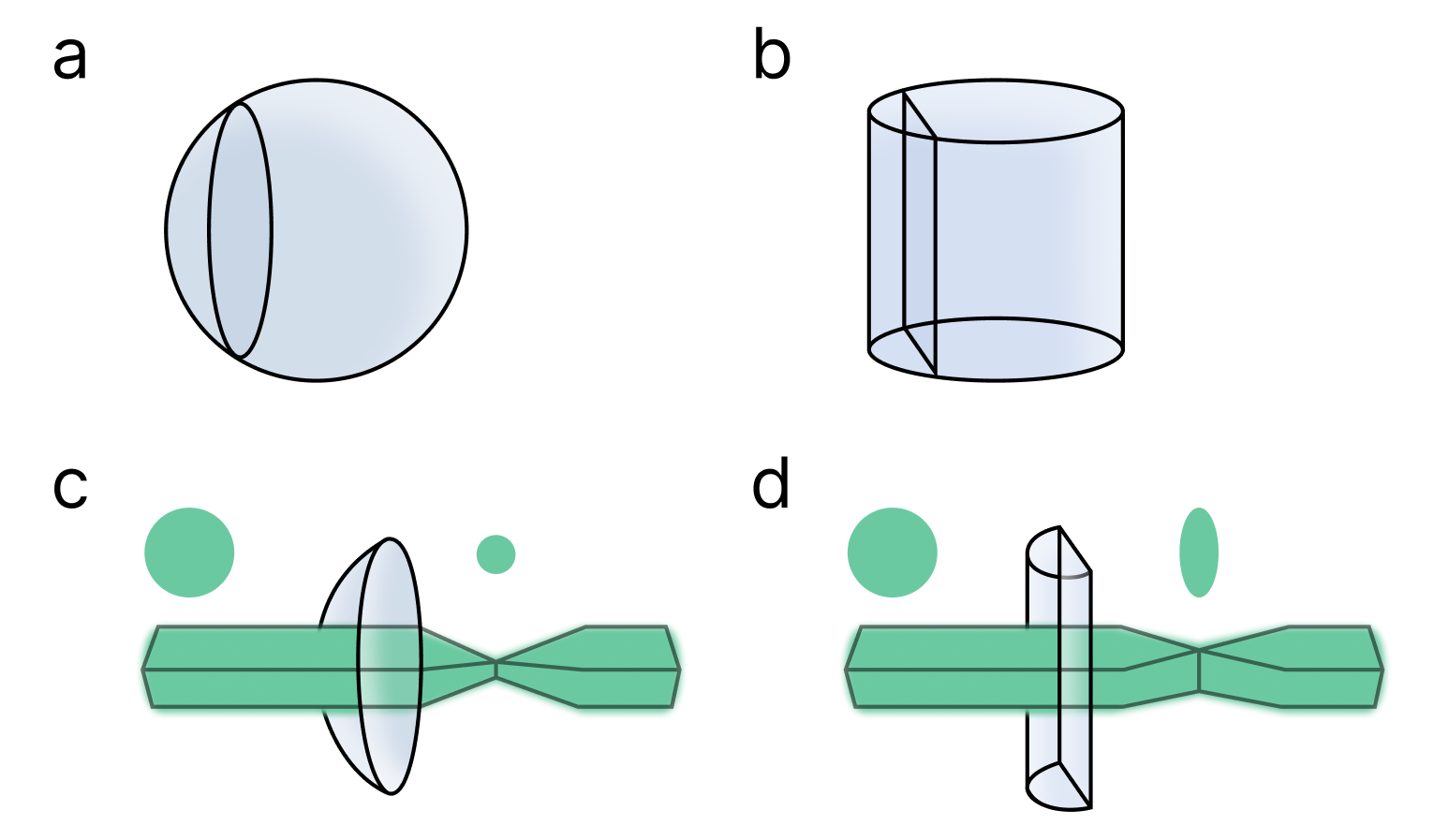}
\caption{\label{fig:sphVScyl}Diagram of a convex spherical (a) and a convex cylindrical (b) lens. Spherical lenses will focus light symmetrically (c) while cylindrical lenses will focus light asymmetrically (d).}
\end{figure}

Cylindrical lenses produce an anisotropic light-sheet. Unlike spherical lenses (Fig.~\ref{fig:sphVScyl}a), which interact with light symmetrically (Fig.~\ref{fig:sphVScyl}c), cylindrical lenses (Fig.~\ref{fig:sphVScyl}b) converge or diverge light asymmetrically depending on orientation (Fig.~\ref{fig:sphVScyl}d).\cite{goldsmith_gaussian_1986,shi_gaussian_2014} One cylindrical lens can generate a triangular sheet along the optical axis, while a pair of either concave or convex lenses oriented orthogonally are needed to produce a diverging rectangular light-sheet.\cite{adams_cylinder_2018} In general, SPIM systems are less expensive and simpler to construct than their DSLM counterparts. However, SPIM light-sheets may exhibit intensity variation based on the beam profile and detection objective FOV.

A virtual light-sheet is generated by swiftly scanning an isotropic beam laterally to simulate a planar light-sheet. This method in DSLM requires minimal prior manipulation, resulting in fewer optical aberrations.\cite{keller_reconstruction_2008} An f$\theta$ lens is employed to convert the scan range of the beam into a vertical array of parallel beams.\cite{keller_digital_2010} The control of the scanning range theoretically enables variation in sheet thickness. Additionally, DSLM ensures uniform illumination across the sample sections, unlike SPIM.\cite{keller_reconstruction_2008,olarte_light-sheet_2018} However, the need for rapid beam movement entails the use of electronic scanning mirrors, escalating system cost and complexity.

\subsubsection{\label{sec:Math}Theoretical calculations to determine the sheet thickness}
The equations presented here assume a Gaussian beam profile. To calculate the beam diameter of the sheet, Eq.~(\ref{eq:GBss}) is redefined considering the specific lens and laser beam quality. The redefined equation incorporates additional variables: $f$ is the lens focal length, $\lambda$ the excitation wavelength, $M^{2}$ the laser beam quality, and $\omega(z)$ a spreading function
\begin{eqnarray}
2\omega_{z} = \frac{4f\lambda M^2}{\pi\omega(z)}.
\label{eq:spotsize}
\end{eqnarray}
In Eq.~(\ref{eq:spotsize}), the beam diameter is defined as $2\omega_{z}$ rather than $2\omega_{0g}$ to adjust for optics not precisely positioned at the focal point of the preceding component. The spreading of the beam waist is determined using $\omega(z)$ to accurately calculate subsequent beam diameters as the optical system progresses
\begin{eqnarray}
\omega(z) = \omega_{0g}\sqrt{1+(\frac{z}{z_{Rg}})^2}.
\label{eq:beamsize}
\end{eqnarray}
The parameter $z$ is the distance of the lens from the focal point of the preceding lens. This approach, applicable in traditional microscopy, enables the determination of the excitation laser diameter along the optical path. In DSLM, which generates a symmetric excitation beam akin to traditional microscopy, Eqs.~(\ref{eq:spotsize}) and (\ref{eq:beamsize}) are utilized once each. However, in a SPIM configuration using cylindrical lenses to produce an asymmetric beam, these equations must be applied separately to ascertain beam diameters in both x- and y-directions.

\subsection{\label{sec:Objectives}Objective lenses will dictate the image quality and LSM geometry}
The objective lens serves as a crucial element in an LSM, impacting various components and capabilities of the microscope. The magnification of the objective determines the FOV achievable for imaging the sample as well as the resolving power, which is especially important in super-resolution imaging. Moreover, magnification influences the focal length of the objectives, with higher magnifications corresponding to shorter focal lengths, as described in Eq.~(\ref{eq:ObjFL})
\begin{eqnarray}
f_{obj} = \frac{f_{tl}}{M},
\label{eq:ObjFL}
\end{eqnarray}
where $f_{tl}$ is the focal length of the tube lens and $M$ the magnification of the objective.
This relationship bears significance when deciding on an objective orientation to ensure the sample is within the working distance (WD), \textit{i.e.}, the distance from the front lens of the objective to the focal point. For super-resolution imaging, the WD typically ranges from 0.01 to 10 mm,\cite{choi_optical_2018,mcconnell_novel_2016,birk_super-resolution_2017,galland_3d_2015} posing challenges due to the limited space between the sample and the lens, especially with non-standard objective orientations. 

The resolving power of the objective is determined by the numerical aperture (NA), which is dependent on the refractive index $n$ of the imaging media and the half-angle $\theta$ of the cone of light entering the lens (Eq.~(\ref{eq:NA}))
\begin{eqnarray}
NA = n\sin\theta.
\label{eq:NA}
\end{eqnarray}
A higher NA yields greater lateral ($R_{x,y}$) and axial ($R_{z}$) resolution, allowing for finer feature discrimination, albeit achievable only when imaging through specific media 
\begin{eqnarray}
R_{x,y} = \frac{\lambda}{2NA},
\label{eq:NAresLat}
\end{eqnarray}
\begin{eqnarray}
R_{z} = \frac{2\lambda}{NA^2}.
\label{eq:NAresAx}
\end{eqnarray}

Commercial objectives exhibit variations in properties such as magnification and NA, greatly affecting resolution capabilities. For instance, comparing two objectives with identical magnification but differing NA values illustrates significant differences in lateral and axial resolutions. An example comparison of objectives of NA$_{1}$ = 1.5 and NA$_{2}$ = 0.65 with an emission wavelength of 532 nm, the resolutions would be $R_{x,y}$ $\approx$ 177 nm and $R_{z}$ $\approx$ 473 nm for objective one, and $R_{x,y}$ $\approx$  409 nm and $R_{z}$ $\approx$ 2,518 nm for objective two. Finding a balance between magnification, WD, and NA is crucial in designing an optical set-up. The following subsections delve into available immersion media, objective orientations, and strategies for selecting appropriate options that will work well in the desired experimental parameters.

\subsubsection{\label{sec:Media}Immersion media is dependent on sample}
The main immersion media for light microscopy are oil, air, and water. The application of each media to samples like zebrafish, cells, and diffusing proteins, are discussed here considering what conditions will necessitate the use of each immersion medium. The NA of an objective is also influenced by the refractive index of the medium (Eq.~(\ref{eq:NA})). A comparison of NAs for commercially available objectives with each immersion medium is presented in Table~\ref{tab:ObjComp}.

\begin{table}
\caption{\label{tab:ObjComp}Comparison of the NA and resolution of commercially available air, oil, and water objectives.}
\begin{ruledtabular}
\begin{tabular}{lcr}
Immersion Media&NA Range\footnote{https://www.olympus-lifescience.com/en/objective-finder/}&Resolution Range\footnote{$\mu$m, assuming $\lambda_{EM}$ = 575 nm}\\
\hline
Oil&0.5 -- 1.5&0.192 -- 0.575\\
Air&0.04 -- 0.95&0.303 -- 7.19\\
Water&0.3 -- 1.2&0.240 -- 0.958\\
\end{tabular}
\end{ruledtabular}
\end{table}

Oil, with its high refractive index ($\sim$1.52),\cite{diaspro_influence_2002,hell_aberrations_1993} is often used in super-resolution imaging due to its compatibility with high NA objectives. This medium is commonly employed in inverted microscopes with samples mounted on glass with a similar refractive index (1.518).\cite{gibson_experimental_1991,fouquet_improving_2015,besseling_methods_2015} Oil immersion objectives require direct objective/oil and oil/glass interfaces, limiting usage primarily to one-objective set-ups,\cite{monge_neria_single-molecule_2023,theer_spim_2016}  making oil suitable for all three experimental examples.

Air offers versatility in objective orientation but has the lowest refractive index (1.00), resulting in lower NA objectives and limited application in nanoscale super-resolution tasks such as sub-cellular imaging or tracking protein diffusion. Samples with larger micron to millimeter features, such as zebrafish,\cite{castranova_long-term_2022} are well-suited for air objectives. Although not commonly used in LSM for biophysical imaging, given samples are commonly in aqueous environments, air objectives are simpler to implement if super-resolution is not required. If air objectives are used, they are regularly paired with a medium other than air.\cite{keller_vivo_2013,ding_light-sheet_2017}

Water, while less versatile than air, provides a more adaptable geometry than oil. Proper index matching still necessitates an aqueous medium, but water objectives are not restricted to one-objective set-ups. With a refractive index of 1.33, water objectives offer a decent range of NAs and are either water immersion or water dipping. Water immersion objectives are designed to be used with a drop of water placed between a glass coverslip and the objective lens. Water dipping objectives do not use a coverslip, and instead are meant to have the front lens either dipped or submerged into the sample medium itself. Water objectives can utilize various aqueous media, such as buffer solutions, offering flexibility. Given that both single cells and diffusing proteins typically require a buffer solution, water serves as an ideal medium for these experiments.\cite{shang_nanoparticle_2014,ma_measuring_2016} Overall, water immersion is a compromise between the versatility of air and the higher resolution capabilities of oil.

\subsubsection{\label{sec:Orientation}Objective orientation is dependent on immersion media}
\begin{figure}
\includegraphics[width=0.5\textwidth]{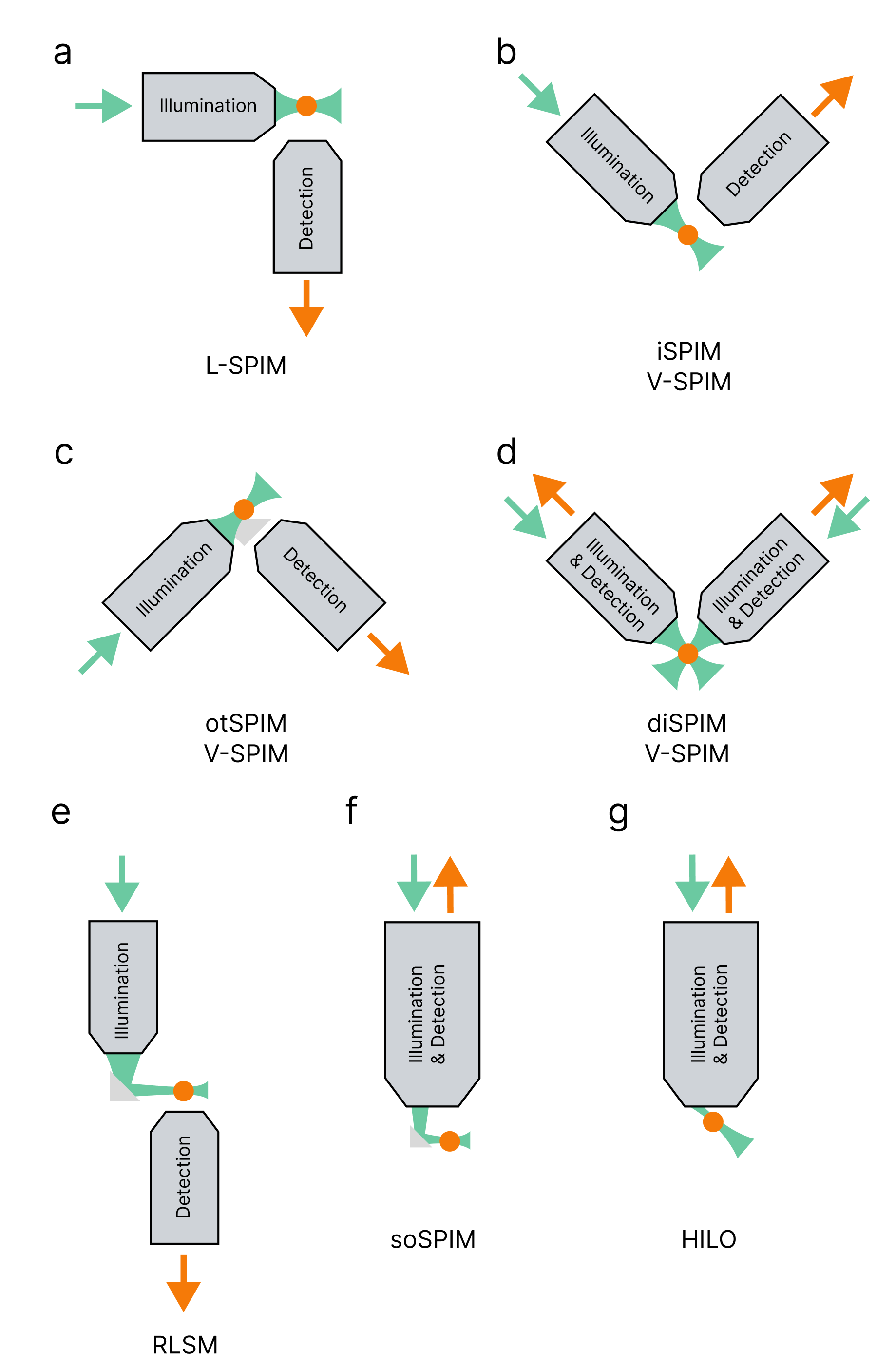}
\caption{\label{fig:objorientation} Diagram of LSM objective orientation options. L-selective-plane
illumination microscopy (SPIM, a) is the easiest two-objective orientation to implement. V-SPIM has several options for objective orientations such as inverted SPIM (iSPIM, b), open-top SPIM (otSPIM, c), and dual-illuminated iSPIM (diSPIM, d). Reflective LSM (RLSM, e) allows for a more straightforward mounting arrangement. Single-objective LSM can be achieved using single-objective SPIM (soSPIM, f) or highly inclined and laminated optical sheet (HILO, g) orientations. Green and orange arrows indicate the directions of the excitation beam and resulting emission. Note: This figure is not comprehensive of all geometries. See Refs.~\onlinecite{girkin_light-sheet_2018,chatterjee_recent_2018,kafian_light-sheet_2023} for a more in-depth discussion on objective geometry options.}
\end{figure}

In most LSM set-ups, the objective geometry consists of two separate optical paths, resulting in an uncoupled illumination and detection. Typically, one objective forms the light-sheet while the other collects the emitted light for detection (Fig.~\ref{fig:objorientation}a-e). It is crucial to consider the WD of both objectives to ensure the confocal parameter (Eq.~(\ref{eq:GBdof})) of the illumination objective aligns with the focal plane of the detection objective. Although using identical objectives facilitates objective housing and WD matching,\cite{capoulade_quantitative_2011,schmid_high-speed_2013} objectives with different magnifications, numerical apertures (NAs), and/or WDs can still be utilized.\cite{ritter_light_2010,keller_reconstruction_2008} The choice depends partly on sample size and the physical gap between the objectives, which is often limited, requiring both objectives be designed for the same immersion media, such as air/air or water/water. However, some geometries do allow for air/water,\cite{tomer_quantitative_2012,krieger_dual-color_2014} air/oil,\cite{migliori_light_2018} or oil/water\cite{zhao_spatial_2014} mismatches in media. It is also worth noting that two-objective orientations are not typically designed to work with oil immersion media, but some can if a mixed media set-up is used.

Some LSMs feature a single optical path where illumination and detection are coupled through a single objective (Fig.~\ref{fig:objorientation}f,g).\cite{galland_3d_2015,meddens_single_2016,monge_neria_single-molecule_2023,saliba_whole-cell_2023} These one-objective configurations are more conventional and easily integrated into standard inverted microscope bodies. Since there is only one objective, immersion media and WD matching are not concerns, allowing for the use of any immersion media when designing a one-objective LSM system. Less common LSM set-ups use three or more optical paths,\cite{huisken_even_2007,royer_practical_2018,krzic_multiview_2012,santi_thin-sheet_2009} but these are beyond the scope of this tutorial. Here, we present one- and two-objective orientations, starting with the more complex two-objective options.

The standard light-sheet configuration, L-SPIM, features orthogonal optical paths, with one objective forming the light-sheet and the other collecting emitted light (Fig.~\ref{fig:objorientation}a). First reported in the early 1900s, L-SPIM is the oldest two-objective set-up.\cite{siedentopf_uber_1902} Although originally the sheet was formed with just a cylindrical lens in the illumination path,\cite{huisken_optical_2004,engelbrecht_resolution_2006,greger_basic_2007} modern set-ups often include an objective after the cylindrical lens to further focus the light-sheet.\cite{wohland_single_2010,singh_performance_2013,oyler-yaniv_tnf_2021} L-SPIM is the simplest geometry and is relatively easy to incorporate into an inverted microscope body, with either the detection or illumination objective being freely mounted above the sample stage. This orientation is suitable for analysis ranging from monitoring the mobility of HP1$\alpha$ in cell nuclei\cite{capoulade_quantitative_2011} to determining the structure of whole insects.\cite{gualda_3d_2022} Additionally, L-SPIM can accommodate both air and water immersion media, though the latter requires a special sample chamber as discussed in Sec.~\ref{sec:Mounting}.

Another two-objective configuration, V-SPIM, maintains objective orthogonality with a 45$^{\circ}$ orientation relative to the sample (Fig.~\ref{fig:objorientation}b-d). Various V-SPIM orientations, such as inverted SPIM (iSPIM), open-top SPIM (otSPIM), and dual-illuminated iSPIM (diSPIM) are primarily suited for air or water immersion media. However, one thing to consider if choosing any of the V-SPIM configurations is the WD of the objectives which should be long enough to ensure the sample does not contact the lenses themselves.

iSPIM and iSPIM-like geometries are frequently reported two-objective orientations, often due to the prevalence of lattice LSM.\cite{chen_lattice_2014,malivert_active_2022,moore_multi-functional_2021,shi_smart_2024} It may be slightly misleading to those accustomed to traditional microscopy techniques, as the "inversion" in iSPIM refers not to the orientation of the objectives themselves, but rather to the capability of integrating the SPIM method onto an inverted microscope.\cite{wu_inverted_2011} To mount the two objectives onto the microscope, they are suspended above the sample (Fig.~\ref{fig:objorientation}b).\cite{wu_inverted_2011,wang_axial_2022,bissardon_selective_2022} The overall geometry of iSPIM accommodates the use of both air and water as immersion media, with water being most common in super-resolution applications (see discussion in Sec.~\ref{sec:Media}). The reduced distance between the objective and the sample can pose challenges, making iSPIM less ideal for super-resolution imaging of larger samples such as adult zebrafish or cleared tissue samples at least 1 cm in size. However, iSPIM inherently avoids issues like media mismatch or off-axis optical aberrations caused by conventional coverslip mounting, as the image is collected from above, eliminating the need to interact with the sample mount.

otSPIM facilitates easy sample loading and manipulation within the objective gap in an iSPIM set-up. In otSPIM, the objectives are positioned below the sample mount, creating an "open-top" design (Fig.~\ref{fig:objorientation}c) that is easily accessible for a multitude of samples and sample chambers.\cite{mcgorty_open-top_2015,zhu_large-scale_2023} By situating the objectives at a 45$^{\circ}$ angle beneath the sample stage, they no longer interface directly with the sample. However, this tilted interface geometry presents challenges due to media mismatch. An additional refractive optic, such as a liquid-filled prism,\cite{mcgorty_high-na_2017,kim_open-top_2021} solid immersion lens,\cite{glaser_light-sheet_2017,chen_rapid_2019} or solid immersion meniscus lens,\cite{barner_solid_2019,barner_multi-resolution_2020} is required in otSPIM to mitigate off-axis optical aberrations caused by media mismatch. Despite the improvements introduced by the refractive optic, the sample mount itself can introduce astigmatism since the optical axis of the detection objective is tilted. To counter this astigmatism, a single cylindrical lens can be inserted into the imaging path before the detector.\cite{mcgorty_open-top_2015,mcgorty_high-na_2017} Most objectives selected for otSPIM applications are designed for use in either air or water due to the availability of more reasonable WDs in these immersion media. While it is theoretically feasible to use an oil objective when coupled with a custom oil-matching refractive index prism, this option should only be pursued if the WD is physically practical.

diSPIM effectively eliminates artifacts by isotropically exciting and collecting emission. Prior to this, discussed objective geometries involved uncoupled illumination and detection paths. However, with side-on illumination from a single objective dedicated to sheet formation, shadow stripes may occur in images due to sample absorption of the excitation source.\cite{glaser_multidirectional_2018,mayer_attenuation_2018,ricci_fast_2020} To mitigate these artifacts, a diSPIM configuration, akin to iSPIM or otSPIM but with both objectives serving as both illumination and detection (Fig.~\ref{fig:objorientation}d), can be employed.\cite{wu_spatially_2013,kumar_dual-view_2014,schueth_efficient_2023,vladimirov_dual-view_2021} Here, a final image is formed by alternating which objective forms the light-sheet and which one captures the image. When both objectives are identical, the system achieves isotropic resolution and diminishes artifacts caused by sample absorption or scattering.\cite{kumar_dual-view_2014,kumar_using_2016} Given the objective geometry similarities with iSPIM and otSPIM, diSPIM set-ups encounter similar advantages and limitations. Unique to diSPIM, illuminating the sample from both sides can increase photobleaching and also requires an added level of complexity due to the need for dual illumination and detection paths.

Reflective LSM (RLSM) is a two-objective method that does not require the objectives to be orthogonal. Instead, the illumination objective is positioned nearly in-line with the detection objective, albeit with a slight offset (Fig.~\ref{fig:objorientation}e). Consequently, an additional reflective component, like a prism\cite{greiss_single-molecule_2016,wang_reflective_2024} or mirror\cite{zhao_spatial_2014,gebhardt_single-molecule_2013} set at a 45$^{\circ}$ angle, is necessary after the illumination objective to direct the light-sheet orthogonally through the sample, hence the name reflective LSM. RLSM accommodates objectives designed for different immersion media within the same set-up. With its predominantly vertical geometry, RLSM can be seamlessly integrated into a standard inverted microscope, enabling the use of conventional sample mounting techniques. In this configuration, the sample is positioned at the detection objective, which offers the option of employing objectives with higher NA, such as those designed for oil immersion.\cite{gebhardt_single-molecule_2013,jannasch_fast_2022,kuhn_single-molecule_2022} However, the incorporation of a small prism or mirror within the narrower WDs of these objectives introduces additional complexity.

One-objective configurations require supplementary optical components to guide the light-sheet through the sample while simultaneously enabling the objective to collect sample emission. Single-objective SPIM (soSPIM) resembles RLSM in using a micro-fabricated mirror\cite{galland_3d_2015,meddens_single_2016,saliba_whole-cell_2023} or prism,\cite{nelsen_combined_2020,beicker_vertical_2018} set at a 45$^{\circ}$ angle to direct the light-sheet orthogonally through the sample (Fig.~\ref{fig:objorientation}f). soSPIM offers the advantage of requiring only a single objective, facilitating easier integration of the light-sheet into a conventional inverted microscope. However, only using one objective can be limiting since the additional reflective component requires careful sample mounting to ensure the sample is within the FOV of the objective while also allowing the light-sheet to reach the reflective optic before entering the sample. Custom sample holders with embedded reflective optics are commonly used to ensure orthogonal penetration of the light-sheet into the sample.\cite{zagato_microfabricated_2017,coelho_direct-laser_2022,beghin_automated_2022} While theoretically compatible with all immersion media, soSPIM may still encounter refractive index mismatches depending on the chosen sample holder.

Highly inclined and laminated optical sheet (HILO) microscopy, often implemented on total internal reflection fluorescence (TIRF) microscopes (Fig.~\ref{fig:objorientation}g), utilizes a single-objective configuration. HILO employs an illumination beam that enters the objective just below the critical angle, generating an inclined sheet.\cite{konopka_variable-angle_2008,tokunaga_highly_2008,knight_single-molecule_2014} This tilting of light is achieved with a higher NA objective in conjunction with a translation optic, typically already incorporated into a conventional TIRF microscope set-up. The excitation beam is translated somewhere between the edge of the back focal plane of the objective (where TIRF occurs) and the center of the objective (where epifluorescence occurs).\cite{bergmann_super-resolution_2018,kozgunova_versatile_2019,monge_neria_single-molecule_2023} Its easy integration into existing TIRF microscope set-ups commonly found in optics labs renders HILO the most widely adopted light-sheet method in many core facilities.\cite{scalisi_single-molecule_2023} TIRF can theoretically be achieved with any of the previously discussed immersion media, making HILO also a viable option. However, water and oil are better suited to this geometry due to their higher inherent NAs. Since the sample is mounted above the objective without additional optical components limiting its placement, various mounting options are available. The primary limitation of the HILO method is the shallower penetration depth of the light-sheet, reaching up to approximately 10 $\mu$m,\cite{tang_extended_2018} compared to other LSM options compatible with samples ranging from 10 to 1000 $\mu$m thick.\cite{huisken_selective_2009}

\subsection{\label{sec:Mounting}Sample geometry and mounting is based on the objective orientation and sample size}
\begin{figure}
\includegraphics[width=0.5\textwidth]{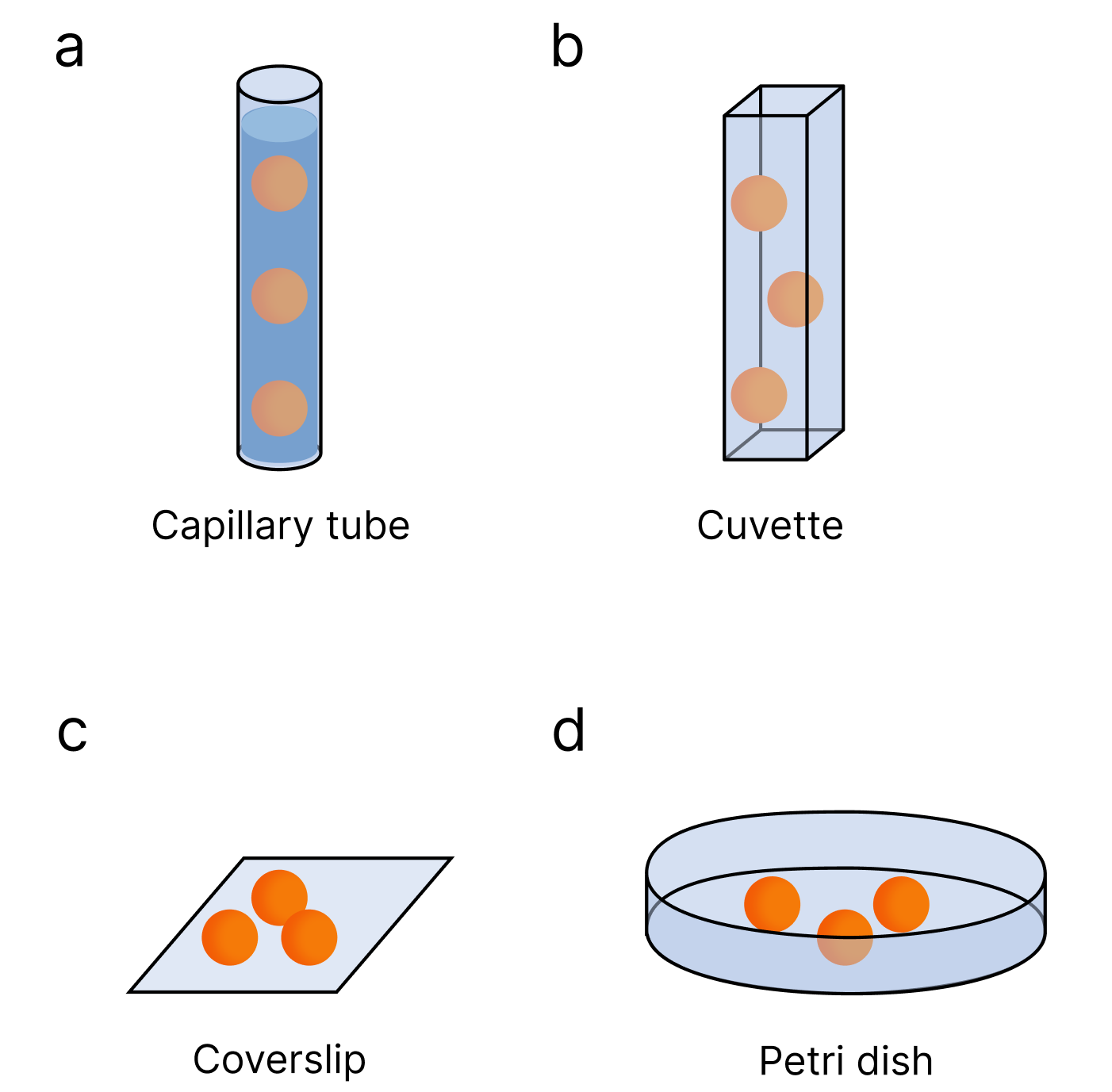}
\caption{\label{fig:sample} Diagram of sample mounting techniques. Capillary tubes (a) have been used to mount samples such as zebrafish embryos. This sample holder relies on index matching between the capillary tube, hydrogel, and media to decrease optical aberrations and is commonly used in water immersion systems. Cuvettes (b) are similar to capillary tubes, but have been used for samples ranging in size from seaweed to cancer cells. These sample holders work well for air and water, but will suffer from optical aberrations due to and index mismatch. Coverslips (c) and Petri dishes (d) are common sample mounting options in conventional microscopy. Both work well for smaller samples such as single cells. Depending on the objective geometry, both coverslips and Petri dishes can be imaged from above or below.}
\end{figure}

LSM requires a level of creativity in how to introduce the sample to the imaging plane, considering both objective orientation and sample size. Minimizing optical aberrations due to media mismatch is also crucial. Fig.~\ref{fig:sample} outlines commonly reported sample mounting techniques. Regardless of the chosen method, it is essential to plan how to acquire a 3D image. Typically, the sample is translated or rotated within the focal plane of the objective(s). Here, we explore a subset of mounting options, focusing on those not tied to specific experiments. Table~\ref{tab:MountComp} includes common immersion media and objective geometries paired with each technique.

\begin{table*}
\caption{\label{tab:MountComp}Comparison of the applicable detection objective immersion media and overall objective geometries for common sample mounting techniques.}
\begin{ruledtabular}
\begin{tabular}{lcr}
Sample Mounting&Immersion Media&Objective Geometry\\
\hline
Capillary&Water&L-SPIM, RLSM\\
Cuvette&Air, Water&L-SPIM, RLSM\\
Coverslip&Air, Water, Oil&iSPIM, otSPIM, diSPIM, RLSM, soSPIM, HILO\\
Petri Dish&Air, Water, Oil&iSPIM, otSPIM, diSPIM, RLSM, soSPIM, HILO\\
\end{tabular}
\end{ruledtabular}
\end{table*}

Embedding techniques are employed for imaging larger samples like zebrafish, involving the placement of the sample within an optically transparent medium encased in a transparent holder (Fig.~\ref{fig:sample}a). In many LSM applications, water serves as the immersion medium, leading to the common use of agarose hydrogel to physically immobilize the sample. Agarose has a high water content ($\geq$ 95\%), resulting in a refractive index (1.335) similar to water (1.33),\cite{fujiwara_agarose-based_2020,engelbrecht_resolution_2006} facilitating imaging through the hydrogel with minimal aberrations. Capillary tubes are often utilized as sample holders in this mounting technique, offering flexibility in their application. Initially, these tubes were used to shape the agarose into a cylindrical form, with a small plunger extruding the gel to align it within the focal plane of the objective.\cite{huisken_optical_2004,pampaloni_third_2007,keller_digital_2011} However, there is a delicate balance in agarose concentration (wt\%) to ensure structural integrity without adversely affecting the sample. Higher concentrations ($\geq$ 1wt\%) may exert compression forces on or restrict the movement of living samples, potentially hindering studies focused on long-term dynamics such as embryo development.\cite{reynaud_light_2008,keller_reconstruction_2008} A modern alternative involves utilizing thin fluorinated ethylene propylene (FEP) capillary tubes instead of silica glass. FEP shares a refractive index of 1.338 with water and agarose, enabling direct sample imaging through the tube.\cite{kaufmann_multilayer_2012,yang_heterogeneities_2022,he_vivo_2023} With no need for hydrogel extrusion, lower agarose concentrations (< 1wt\%) can be used, effectively immobilizing the sample without impeding any dynamic studies.\cite{kaufmann_multilayer_2012} Additionally, to match the refractive index of water, capillary tubes are often suspended within a water-tight chamber, ranging from custom-made\cite{pampaloni_tissue-culture_2014} to commercially available.\cite{jahr_eduspim_2016}

Four-sided cuvettes offer structural stability benefits akin to capillary tubes (Fig.~\ref{fig:sample}b). Typically, cuvettes are treated similarly to capillary tubes, employing a gelling agent to suspend the sample within the chamber. This method accommodates various sample sizes, from seaweed fragments\cite{lichtenberg_light_2017} to breast cancer cells.\cite{kafian_light-sheet_2020} Alternatively, cuvettes can serve as pseudo media baths for samples. Here, cells are cultured on a conventional coverslip, which is affixed to the bottom of the cuvette, followed by filling with the desired media.\cite{gustavsson_3d_2018} While cuvettes provide a straightforward commercially available mounting option, they have inherent drawbacks. Primarily, although optically clear, four-sided cuvettes are commonly made from materials with higher refractive indices than the sample. Most commercially available cuvettes are crafted from materials like quartz, borosilicate glass, or polystyrene, with refractive indices around 1.459, 1.519, and 1.55, respectively, whereas LSM-imaged samples typically have refractive indices ranging from approximately 1.33 to 1.47.\cite{nussbaumer_simple_2005,kaufmann_multilayer_2012,murray_simple_2015} Consequently, this refractive index mismatch leads to spherical aberrations.\cite{williams_practical_2023} To address this, one might consider using objectives designed for higher refractive index media, such as oil. However, oil immersion objectives, as discussed in Sec.~\ref{sec:Media}, have the shortest working distance among immersion media, typically in the range of a few hundred micrometers.\cite{choi_optical_2018} Given that the thinnest wall of a cuvette is 1 mm, oil immersion objectives would not be suitable in this context.

The simplest mounting techniques, familiar to those knowledgeable of conventional microscopy, include glass coverslips (Fig.~\ref{fig:sample}c) and Petri dishes (Fig.~\ref{fig:sample}d). Coverslips, while common, introduce complexity to sample mounting due to objective orientation. When deciding between one- and two-objective orientations, consideration of physical space for the sample is important, with more constraint typically in two-objective geometries (see Sec.~\ref{sec:Orientation}). 
The physical limitations of a coverslip also impact the size of imageable samples. In standard inverted orientations, samples are imaged from below, requiring the light-sheet to pass through the coverslip, leading to having to select an appropriate coverslip thickness. Typically, a commercially available No. 1.5 coverslip with a thickness range of 160 - 190 $\mu$m is suitable. Opting for an objective geometry that images the sample from above affords more flexibility in sample preparation and immersion media. Since the coverslip is not in the imaging path, refractive index mismatches due to the substrate are less likely, and any glass thickness can be chosen. In such cases, oil immersion is not viable, while both water and air are. When using water, maintaining the objective and sample in the media throughout imaging is necessary, often facilitated by a media bath component, ranging from custom fabrication\cite{chen_lattice_2014} to a commercially available Petri dish.\cite{capoulade_quantitative_2011} 

Petri dishes have been used in conventional optical microscopy facilitate the imaging of biological samples under controlled environments. These dishes accommodate various requirements, from cell culturing to serving as a media bath or a combination of both functions.\cite{sapoznik_versatile_2020,capoulade_quantitative_2011} Similar to glass coverslips, Petri dishes offer flexibility in imaging orientations, supporting sample observation from either above or below. When imaging from below, it is optimal to use glass-bottom dishes equipped with a No. 1.5 coverslip window to match an oil refractive index for improved imaging quality. For objectives that image from above, standard plastic Petri dishes are sufficient. However, it is crucial to ensure that the dish dimensions align with the physical constraints of the objectives and samples. The dish diameter should accommodate the objective(s) without making contact, while the dish depth must accommodate the sample and immersion media. Fortunately, there is a wide range of commercially available Petri dish dimensions that fit the requirements of specific LSM objectives and samples.

\subsection{\label{sec:Detection}Cameras are chosen based on the desired sensitivity and timescale of the measurements}
Two-dimensional scientific cameras are used to detect the collected emission from the sample. electron multiplying charge-coupled device (EMCCD) or scientific complementary metal-oxide-semiconductor (sCMOS) cameras are the two options for nanoscale imaging due to their high (up to 95\%) quantum efficiencies. Cameras are selected based on the noise, sensitivity, and frame rate capabilities.

EMCCD cameras find extensive application in various low-light imaging scenarios, spanning from biological samples\cite{chizhik_super-resolution_2016,song_situ_2023} to single-molecule measurements\cite{verma_single-molecule_2016,blanquer_relocating_2020} to astronomy,\cite{jaimes_exploring_2016,gili_measurements_2020} where high sensitivity is crucial for accurate imaging. The inherent high sensitivity of EMCCD cameras stems from an on-chip multiplication gain mechanism that enables the detection of single-photon level intensity. However, this gain mechanism introduces additional noise, potentially lowering the overall signal-to-noise ratio (SNR) and image quality.

To enhance sensitivity, EMCCD cameras typically feature larger pixels, typically around 13 $\mu$m. While this increases photo-sensitivity, it comes at the expense of resolution. The challenge lies in selecting an appropriate magnification objective to ensure that the image pixel size satisfies the Nyquist sampling criterion, which dictates that the pixel size should be at least half the size of the object being resolved.\cite{nyquist_certain_1924} In fluorescence microscopy, the "object" size is determined by the diffraction limit of light, approximated by Eq.~(\ref{eq:DiffLim})
\begin{eqnarray}
d = \frac{\lambda}{2 NA},
\label{eq:DiffLim}
\end{eqnarray}
where $d$ is the diffraction limit, $\lambda$ the emission wavelength, and $NA$ the numerical aperture of the detection objective. In systems with an emission wavelength at or above 500 nm, and assuming an NA of 1, the correlating "object" size is at least 250 nm. Therefore, to meet the Nyquist criterion, the pixel size should ideally be close to or less than 100 nm. For an ideal system, the required magnification can be calculated using Eq.~(\ref{eq:PxSize})
\begin{eqnarray}
\frac{P_{P}}{M} = P_{I},
\label{eq:PxSize}
\end{eqnarray}
where the physical pixel size ($P_{P}$) is divided by the objective magnification ($M$) to obtain the image pixel size ($P_{I}$). For most EMCCD cameras with a $P_{P}$ of 13 $\mu$m, achieving a $P_{I}$ of 100 nm needs a magnification of 130X. While this magnification may seem high, it is feasible for many super-resolution applications and can be achieved with a 100X objective supplemented by additional magnification optics in the detection path.

Due to the increased light-sensitivity, EMCCD cameras are limited in achievable frame rates. The on-chip multiplication process restricts the speed at which the camera can read and clear the signal.\cite{qiao_method_2021} While imaging at full-chip, these cameras typically operate at 61 fps, but can reach speeds as high as 4,000 fps with a smaller FOV as well as using pixel binning.\cite{andor_ixon_nodate} Accordingly, EMCCD cameras are well-suited for imaging static or slow-moving samples with low-emission, where resolution below 130 nm is not essential.

sCMOS cameras, a newer technology compared to their EMCCD counterparts, are increasingly employed in biological imaging\cite{mandracchia_fast_2020,sadoine_designs_2021} and dynamic studies.\cite{basumatary_temporally_2023,fujiwara_development_2023} While they are less photosensitive than EMCCDs due to the absence of a gain mechanism, sCMOS cameras exhibit lower overall read noise. However, sCMOS cameras do suffer from pixel-dependent noise, which is difficult to correct.\cite{liu_scmos_2017} The primary advantages of sCMOS cameras lie in their higher inherent resolution and fast achievable frame rates. The enhanced resolution is attributed to their smaller average pixel size, approximately 5 $\mu$m. Referring back to Eq.~(\ref{eq:PxSize}), and maintaining the same assumptions, a resolution of 100 nm can be easily attained with a 50X magnification. Considering typical magnifications of 40X to 100X are used in super-resolution microscopy set-ups, this is not an unreasonable magnification to achieve without additional optics following the objective. The faster frame rates achievable with sCMOS cameras are primarily due to their clearing mode. Unlike EMCCDs, which clear the entire chip simultaneously, sCMOS cameras clear column by column, enabling full-chip frame rates exceeding 100 fps.\cite{fowler_55mpixel_2010} These cameras often come equipped with various settings optimized for SNR, speed, or dynamic range. By imaging a smaller FOV, frame rates of up to 10,000 fps can be reached using a speed setting.\cite{muschol_activity-dependent_2003} However, this increased speed comes at the expense of sensitivity. Consequently, sCMOS cameras are best suited for imaging samples with high emission efficiencies, relatively fast movement typical of dynamic studies, and resolution requirements below 130 nm.

\section{\label{sec:Alignment}Alignment and calibration procedures for LSM}
\subsection{\label{sec:OurLSM}Design elements chosen for our home-built LSM}
\begin{figure*}
\includegraphics[width=0.99\textwidth]{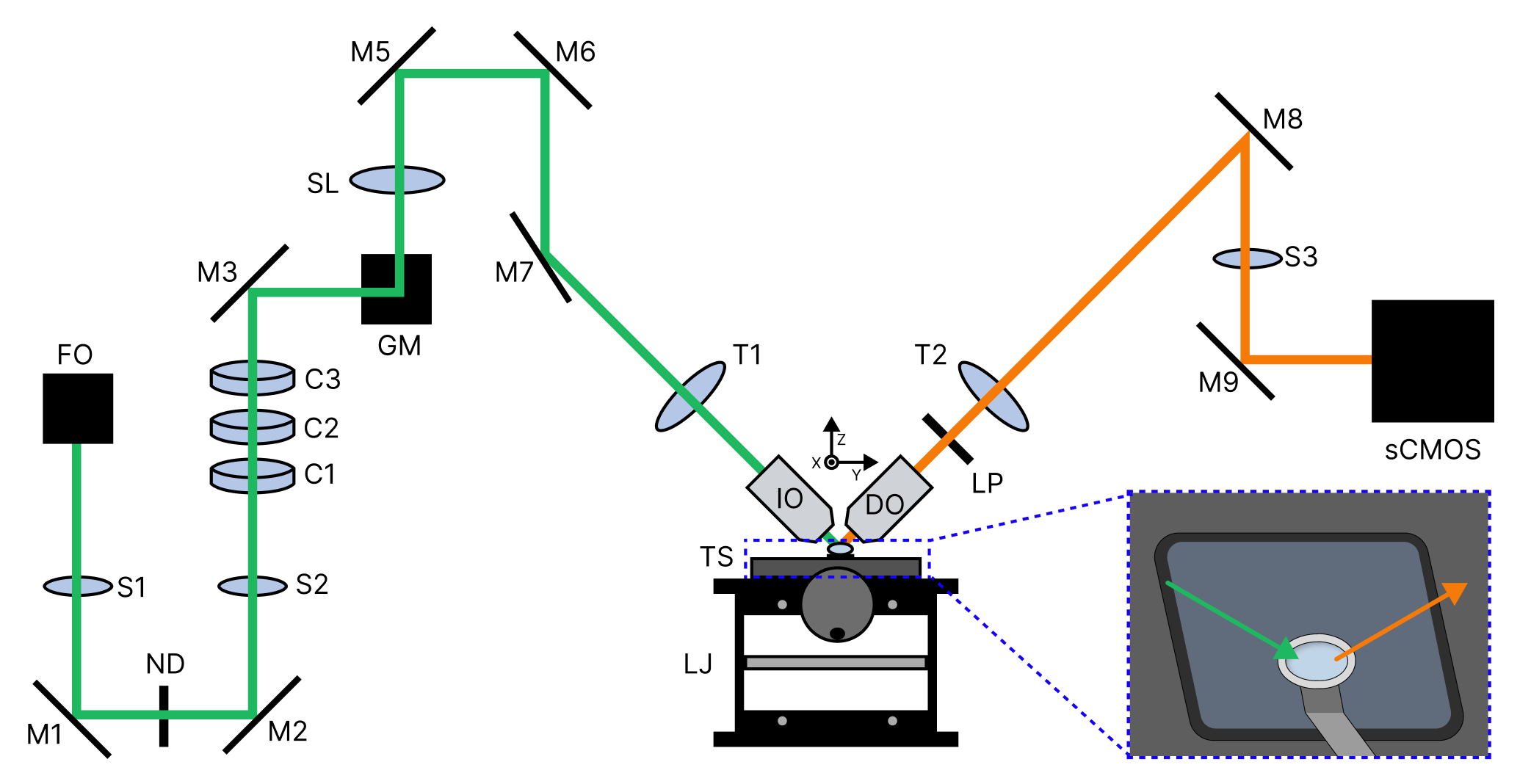}
\caption{\label{fig:beampath} Simplified beam path of our light-sheet microscope. A linearly polarized 1.14 mm $\varnothing$ 532 nm Gaussian beam is introduced into the optical path via a fiber optic coupling (FO), compressed symmetrically by a pair of convex spherical lenses (S1,2), stretched along the y-axis by a pair of convex cylindrical lenses (C1,3), and then compressed along the x-axis by a convex cylindrical lens (C2) to form a sheet. The sheet is directed through an f$\theta$ lens (SL) via two galvanometer mirrors (GM), passed through a tube lens (T1), and enters the back pupil of the illumination objective (IO). The light-sheet excites the fluorophores within a sample and the emission is imaged with a detection objective (DO), tube lens (T2), relay lens (S3), and sCMOS camera. The intensity of the beam is varied using a neutral density filter wheel (ND) along the illumination path. To ensure only the sample emission reaches the camera, a 550 nm long-pass filter (LP) is placed along the detection path. Here, green denotes the excitation path while orange denotes the detection path. For our imaging procedure, the sample is translated along the z-axis through the light-sheet via a piezoelectric stage mounted on top of an xy-translation stage (TS). For course adjustments along the z-axis, we use a lab jack (LJ) custom fit with the sample stage. Inset: The 10 mm $\varnothing$ sample (light blue circle) is mounted on a custom stainless steel arm suspended within an aqueous media bath. The arrows represent the locations of the objectives relative to the sample.}
\end{figure*}

The upcoming sections on microscope alignment and calibration are intricately linked to the specific LSM system in use. To provide a practical example, we will first outline the set-up of our home-built system (Figs.~\ref{fig:beampath}, S1 and Table~S1). As previously mentioned in Sec.~\ref{sec:Intro}, our microscope aims to capture features beyond the cell, particularly monitoring protein diffusion within the ECM.\cite{kisley_characterization_2015,yoshida_computationally-efficient_2021,chatterjee_spatially_2023,antarasen_cross-correlation_2024} Consequently, considerations for resolving nanoscale features and dynamics of 10 $\mu$m$^2/s$ diffusion \textit{in situ} dictate our choices regarding beam profile, immersion media, objective orientation, sample mounting, and camera selection.

For excitation in our LSM, we opt for a simple Gaussian beam profile for its ease of implementation. Since we focus on nanoscale imaging rather than deep tissue penetration, complex beam profiles like Bessel or Airy are unnecessary, as we will not be working with larger mm-scale tissue samples. Our sheet formation involves a combination of cylindrical lenses (Fig.~\ref{fig:beampath}, C1-3) and a scanning galvanometer mirror system (Fig.~\ref{fig:beampath}, GM). A galvanometer mirror consists of a lightweight mirror attached to a coil within a magnetic field. By varying the current through the coil, the mirror can rapidly and precisely scan the laser beam across a surface in one dimension per mirror. This hybrid approach is meant to mitigate shadow formation issues inherent in Gaussian beams. Using Eqs.~(\ref{eq:spotsize})-(\ref{eq:beamsize}) from Sec.~\ref{sec:Math}, we calculate theoretical beam dimensions based on the lenses and beam path of our microscope, resulting in dimensions of 11.23 $\mu$m x 4.24 $\mu$m. Later, in Sec.~\ref{sec:CalSheet}, we will describe our method for determining the experimental beam dimensions and compare them to these theoretical values.

We opt for water as our immersion media due to its refractive index compatibility with most biological environments and their mimics. This choice allows flexibility in objective geometry, as discussed in Sec.~\ref{sec:Orientation}. Considering our intended samples, such as proteins diffusing within an ECM-like matrix, we employ identical 40X water dipping objectives (refer to Table~S1) oriented in an iSPIM configuration. However, the limited space between objectives in the iSPIM orientation requires careful sample mounting. For the objectives we chose, the WD is considered relatively long at 3.3 mm. This WD does not afford a very large objective gap ($\sim$12 mm at the base), resulting in the footprint of the sample having to be smaller. Additionally, as depicted in Fig.~\ref{fig:objorientation}b, the objective gap is triangular. Therefore, we designed a method (detailed in Methods in SI) to create a $\sim$3 mm high domed sample on a 10 mm diameter glass coverslip, fitting within the narrow objective gap while still reaching the focal plane. The sample is immobilized by covalently bonding the ECM-like matrix to the glass (Methods in SI) to minimize sample drift and blurring during imaging, and then mounted on a custom arm within a media bath (Fig.~\ref{fig:beampath} inset and CAD files S1 and S2).

To acquire 3D datasets, we move the sample through the light-sheet while collecting emission intensity. Sample movement is achieved by mounting the media bath to a xy-translation stage (Fig.~\ref{fig:beampath}, TS) and mounting the sample arm onto a piezoelectric stage for vertical translation. With an NA of 0.8 and an emission wavelength of ~575 nm, we anticipate sub-micrometer axial resolution (360 nm) and micrometer lateral resolution (1.8 $\mu$m) (Eqs.~(\ref{eq:NAresLat}),(\ref{eq:NAresAx})). Given our focus on fast dynamics, we have chosen a sCMOS camera over an EMCCD for its higher frame rate capabilities. Additionally, the higher resolution of the sCMOS suits imaging nanoscale structures. For further specific component descriptions and dimensions, refer to Table~S1 and CAD files S1 and S2. 

\subsection{\label{sec:AlignStand}Standards for alignment should be detectable by eye}
Aligning a microscope does not require a specific standard until the detection path. It is recommended to start with a highly fluorescent bulk bead or dye solution, typically on the order of $\mu$M, allowing the beam to be visible to the naked eye (Fig.~S2), which aids in troubleshooting. Moreover, using a bulk sample eliminates the possibility of imaging molecular diffusion. Once the sample emission reaches the camera, a less concentrated solution should be used. This lower concentration offers a clearer understanding of alignment by revealing any aberrations of individual emitters in the camera. To counter potential blurring from diffusion, suspending the particles within a gel with a refractive index similar to that of the solvent can be helpful. For our standard, we utilized a 2.0 $\mu$m bead solution diluted to 1:10 of the stock (Methods in SI).

\subsection{\label{sec:AlignProcess}Alignment is an iterative process }
The alignment process may appear daunting initially, but following a systematic approach can simplify it considerably. In this section, we will walk through our LSM alignment process and provide useful tips for other set-ups.

To begin alignment, map out the optical mount locations on the breadboard. The spacing between lenses, determined by their functions and focal lengths, should be calculated beforehand (Sec.~\ref{sec:Math}). Since this spacing may not align with the inherent spacing of the breadboard, incorporating rails or movable post mounts can be advantageous. Additionally, decide whether optics will be mounted horizontally or vertically. For our LSM, we opted for vertical mounting to accommodate the iSPIM orientation of the objectives, which was achieved using a custom "U"-shaped breadboard (Base Lab Tools, Fig.~S1).

After mapping the path, attach the optical mounts onto the breadboard. Ensure consistency in the center of each mount component to aid lens alignment later. Although components of the same shape and size are helpful, this is not always feasible, so adjustments may be needed for those with different dimensions. Using post mounts with adjustable heights (\textit{e.g.}, ThorLabs, PH2) and a non-reflective ruler (\textit{e.g.}, ThorLabs, BHM3) can be beneficial in ensuring each component is centered at the same height. However, for vertical mounting, there is potential for misalignment due to gravity and the post is only secured by a single side screw within the mount. The impact of gravity can be mitigated by using posts that have been machined to the desired height and secured directly to the breadboard. During mounting, ensure all mirror mounts are set to 45°, while lens mounts are either at 90° or 180° relative to each other for easy beam path tracing. While some LSM geometries may require deviations, maintaining this rule for most components simplifies alignment with irises.

Once optical mounts are in place, ensure the laser passes through the center of each mount to facilitate overall microscope alignment and prevent edge aberrations. Start with the laser at its lowest power setting, using fluorescent targets (\textit{e.g.}, ThorLabs, VRC2RMS) along the optical path for beam location identification. If this value is greater than $\sim$5 mW, neutral density filters should be used to decrease the overall laser intensity. Protective eye-wear (\textit{e.g.}, ThorLabs, LG12) should also be worn during this process.\cite{hill_jr_chapter_2010} In this first alignment iteration, a multi-iris system with optical post collars (\textit{e.g.}, ThorLabs, R2) to fix the iris height can be used to ensure the laser beam is centered on each mounting component. Mirrors should be used to direct the beam through each iris until it reaches the objective.

In subsequent alignment iterations, introduce optics into the path gradually to avoid deviations caused by a slight tilt in the optic. If all the lenses are introduced at once, isolating the exact optic(s) causing the deviation will be difficult. In an SPIM set-up, a two-iris system can aid alignment until the sheet formation stage at the cylindrical lenses. The two-iris system can generally be used throughout the illumination path of a DSLM since the beam remains circular. For aligning sections with asymmetric beam shapes, a rail system may be necessary. If the beam is aligned prior to entering the first lens, a rail system should allow the beam to maintain its trajectory through the other lenses, which is advantageous when the lens placements do not afford enough room for an iris to be mounted. After the cylindrical lenses, the alignment of the beam can be determined using the fluorescent targets. Using the previous mirror within the optical path, it is possible to direct the beam to the center of the next mirror. This method was used from the cylindrical lenses (Fig.~\ref{fig:beampath}, C1-3) to the f$\theta$ lens (Fig.~\ref{fig:beampath}, SL) and then again from the f$\theta$ to the illumination tube lens (Fig.~\ref{fig:beampath}, T1). The illumination tube lens and illumination objective (Fig.~\ref{fig:beampath}, IO) should be in line with one another. Alignment typically requires multiple attempts and refinements, but once everything seems correct, confirm the beam exits the objective at a 45° angle relative to the sample stage using the highly fluorescent alignment standard (Sec.~\ref{sec:AlignStand}).

The detection path alignment, theoretically simpler than illumination, involves centering the emission collected by the detection objective (Fig.~\ref{fig:beampath}, DO) on the detection tube lens (Fig.~\ref{fig:beampath}, T2). Use a mirror to direct emission light through a relay lens (Fig.~\ref{fig:beampath}, S3), ensuring no warping of the image with the aid of an iris. After the relay lens is a mirror used to direct the light into the camera. Once emission reaches the camera, fine-tune the sheet orientation using a lower concentration of the alignment standard solution, where individual particles can be seen, to better diagnose any aberrations caused by slight misalignments.

\subsection{\label{sec:Calibration}Calibrations for nanoscale imaging require single particle and molecule samples}
Once the microscope is well aligned, calibration is essential to begin data collection. Understanding the FOV based on the beam shape and size, along with the precise pixel size of the set-up, is crucial for orienting within the sample. Additionally, it is important to acknowledge that there may be image skewing inherent to 3D imaging. Before reconstructing the structure of the sample, identifying and accounting for this offset is necessary.

\subsubsection{\label{sec:CalStand}Standards for calibration should be highly emissive beads of known size}
Our standard calibration sample differs from the alignment sample to enable visualization of individual fixed particles for accurate determination of the microscope pixel size and inherent axial skewing in LSM. To accomplish this, we suspended beads in a 2 wt\% agarose hydrogel. Agarose is well-suited for our water dipping objectives due to its refractive index matching, as discussed in Sec.~\ref{sec:Orientation}.\cite{fujiwara_agarose-based_2020,engelbrecht_resolution_2006} The bead sizes range from 2 $\mu$m to 0.25 $\mu$m, facilitating straightforward determination of pixel size. We selected a bead concentration that ensures sufficient density for statistical analysis of pixel size yet avoids overcrowding that would hinder the distinction between individual beads.

\subsubsection{\label{sec:CalPixel}Pixel size can be measured using beads}
To determine the pixel size of our microscope set-up, we cannot use conventional methods like a USAF target due to our "non-conventional" geometry.\cite{goncalves_estimating_2006,picazo-bueno_design_2022} The flat glass USAF targets are beyond the 3.3 mm WD of the two objectives in our iSPIM set-up, and we are unaware of any 3D AF or NIST imaging target compatible with "non-conventional" geometry LSM. Therefore, we have determined the pixel size by measuring the point spread function (PSF) of highly fluorescent beads of varying sizes, specifically 2 $\mu$m, 1 $\mu$m, and 0.25 $\mu$m beads suspended in a 2 wt\% agarose hydrogel (actual sizes reported by the manufacturer are listed in Table~\ref{tab:PSFbeads}). For this method, we first assume the $P_{I}$ is ideal based on Eq.~(\ref{eq:PxSize}), where $P_{P}$ is 6.5 $\mu$m and $M$ is 40X, resulting in a $P_{I}$ of 0.163 $\mu$m. Using this image pixel size, we estimate the approximate PSF for each selected bead size. 
Once we estimate the number of pixels that each bead size should encompass, we measure the actual pixel size using our localization software and extract the average pixel size of each bead sample, as detailed in Table~\ref{tab:PSFbeads}.\cite{kisley_research_group_cwru_superreskineticslrg_superres_particle_identify_2021} The expected PSF values are statistically within the range of the experimental average PSF values, confirming our alignment in terms of magnification. To determine the actual size of our image pixel for our microscope, we divide the bead size by the average PSF (Table~\ref{tab:PSFbeads}). The average image pixel size across all bead sizes is 0.18 $\pm$ 0.04 $\mu$m, which is larger than the Nyquist criterion due to using a 40X magnification rather than a 60X.
\begin{table*}
\caption{\label{tab:PSFbeads}Comparison of the expected and experimental PSFs for each bead size measured.}
\begin{ruledtabular}
\begin{tabular}{ccccc}
Bead Size\footnote{$\mu$m}&Expected PSF\footnote{px}&Experimental PSF$^{\mathrm{b}}$&Pixel size$^{\mathrm{a}}$&Overall Average Pixel Size$^{\mathrm{a}}$\\
\hline
2.10 $\pm$ 0.09 & 12.9 $\pm$ 0.6 & 11 $\pm$ 2 & 0.18 $\pm$ 0.02 & 0.18 $\pm$ 0.04\\
0.950 $\pm$ 0.024 & 5.83 $\pm$ 0.15 & 5.4 $\pm$ 0.8 & 0.18 $\pm$ 0.03 &\\
0.25 $\pm$ 0.05 & 1.5 $\pm$ 0.3 & 1.4 $\pm$ 0.1 & 0.18 $\pm$ 0.07 &\\
\end{tabular}
\end{ruledtabular}
\end{table*}

\subsubsection{\label{sec:CalOffset}Axial offsets must be quantified for correct 3D reconstruction}
\begin{figure}
\includegraphics[width=0.5\textwidth]{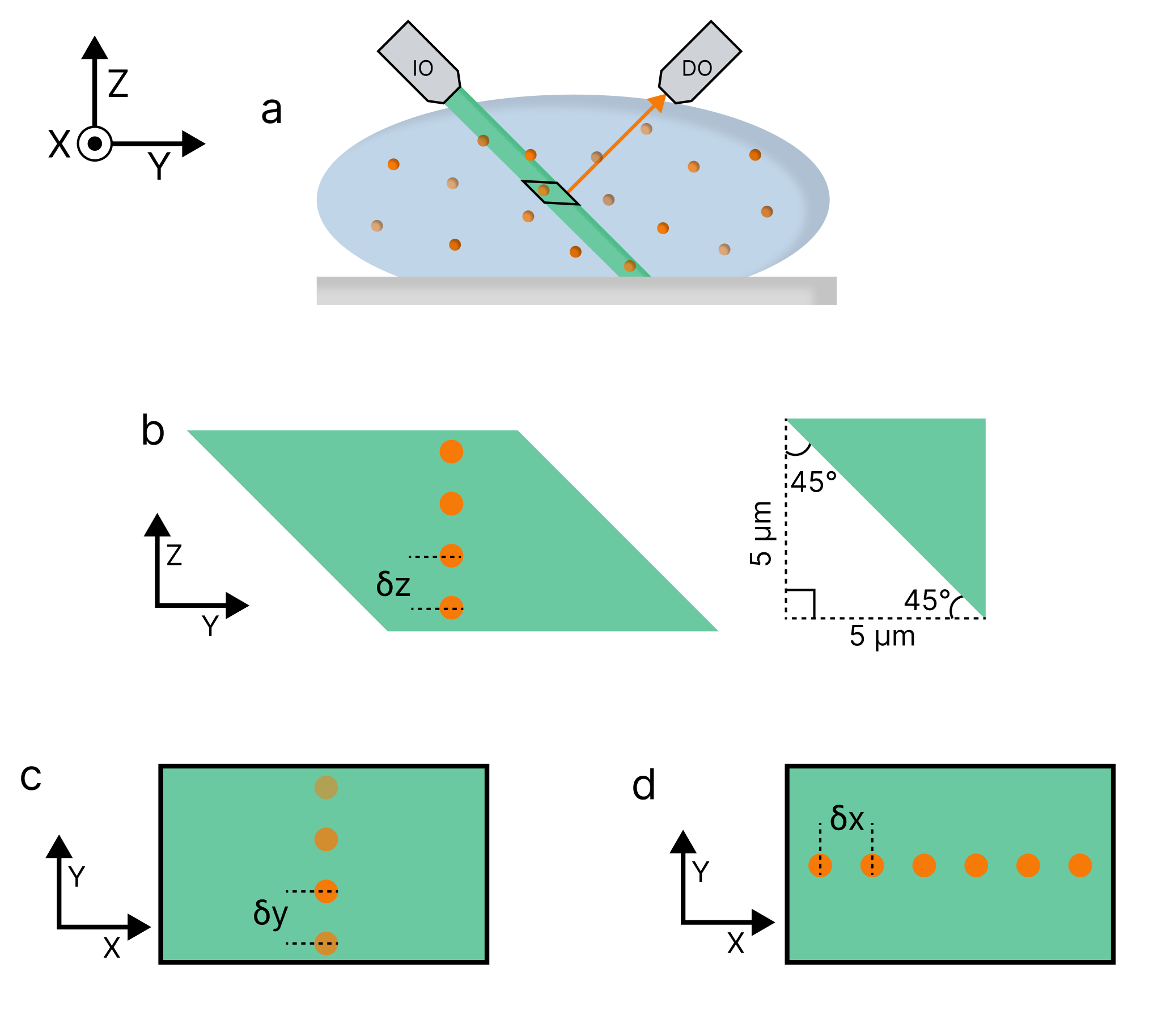}
\caption{\label{fig:geooffset} Generalized diagram of the light-sheet passing through and exciting a sample (a). Notice how the illumination is not perpendicular to the sample itself, leading to a more complicated geometrical offset of fluorophores during the collection and reconstruction process. The portion highlighted in black depicts the section of the sample within the FOV of the detection objective. This subsection will be used to demonstrate the positions of the fluorophores within the FOV relative to the detection objective as the sample is moved within the z-, y-, and x-axes, respectively. Diagram depicting the expected skewed trajectory of the bead standard along the z-axis (b). Here, the sample is vertically within the sheet. The inset demonstrates the expected geometry of the fluorophore offset as it moves along the z-axis. Diagram depicting the expected skewed trajectory of the bead standard along the y-axis (c). Here, the sample is closer to and further from the detection objective. Diagram depicting how the sample stays within the focal plane when moving along the x-axis resulting in no expected offset (d).}
\end{figure}

In our iSPIM set-up, the excitation and detection paths intersect at right angles, offset from the sample itself, resulting in a parallelogram-shaped sample region (Fig.~\ref{fig:geooffset}a). Consequently, when imaging in 3D, we perform three separate calibrations to correct for the skewed data in the z-, y-, and x-axes (Fig.~\ref{fig:geooffset}b-d).

As the sample moves through the light-sheet, the same region is detected multiple times in the FOV, irrespective of its orientation relative to the detection objective. This redundancy is accounted for during 3D image reconstruction. Initially, we focus on calibrating the z-offset, which is the most common type of skewing encountered. This is achieved by moving a 2 $\mu$m bead calibration standard along the z-axis in 5 $\mu$m steps using a piezoelectric stage (Fig.~\ref{fig:geooffset}b). However, experimental analysis (Fig.~S3a) reveals a discrepancy between the expected ($\sim$5 $\mu$m) and observed (2.2 $\pm$ 0.1 $\mu$m) shifts, indicating that the angle of the light-sheet relative to the sample differs from the assumed 45$^{\circ}$ (Calculations in SI).

Although the light-sheet angle minimally impacts z- and x-translations, it significantly affects y-translation and the orientation of the captured image on the camera. Through trigonometric calculations, we determine that the actual angle of the light-sheet relative to the sample is 66$^{\circ}$, not 45$^{\circ}$ as initially assumed. Consequently, the image appears tilted on the camera (Fig.~S4), with shallower emitters appearing closer to the detection objective than deeper ones.

Next, we address the skew along the y-axis, which requires a different correction approach due to geometric considerations (Fig.~\ref{fig:geooffset}c). Using the previously determined angles, we calculate the expected shift along the y-axis (Calculations in SI). Trigonometric calculations yield an expected average shift of approximately 4 $\mu$m, consistent with the experimental observation of 3.8 $\pm$ 0.2 $\mu$m (Fig.~S3b).

Finally, we correct for image skew along the x-axis, which is straightforward since it is parallel to the image plane (Fig.~\ref{fig:geooffset}d). Here, we ensure that the physical shift aligns with the pixel shift. Experimental results confirm that moving the sample 5 $\mu$m along the x-axis corresponds to an average pixel shift of 5.3 $\pm$ 0.2 $\mu$m (Fig.~S3c), as expected.

\subsubsection{\label{sec:CalSheet}Sheet thickness can be measured with a beam profiling camera}
Originally, our aim was to determine the sheet thickness by tracking the duration a bead remains in focus as we move the sample along the z-axis through the sheet. However, this approach proved impractical due to the geometrical factors discussed in Sec.~\ref{sec:CalOffset}, where the initial position of a feature in the fluorescent sample within the sheet affects its observability, rather than the sheet thickness itself. Consequently, we opted to physically measure the beam using tools like a beam profiling camera capable of resolving dimensions down to 20 $\mu$m (\textit{e.g.}, Thorlabs, BC207VIS). 

\begin{figure}
\includegraphics[width=0.5\textwidth]{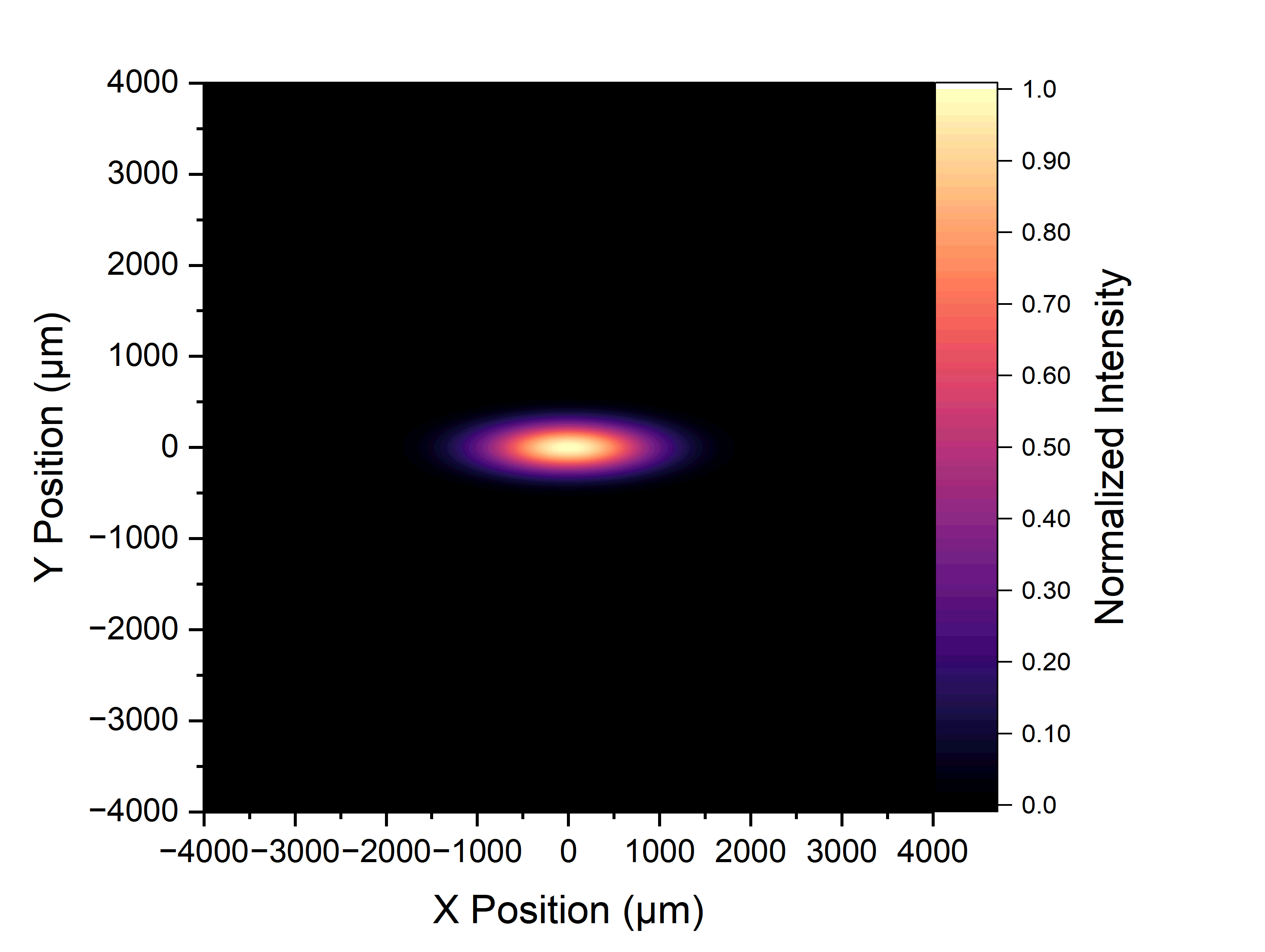}
\caption{\label{fig:physbeam} Beam profile of the light-sheet measured by a profiling camera prior to reaching the illumination objective.}
\end{figure}

To ensure accuracy given the beam size is smaller than the resolution limit of the camera, we measure prior to the objective (Fig.~\ref{fig:physbeam}) and then calculated the final size using the magnification of the objective.
Applying Eq.~(\ref{eq:spotsize}), we determined the experimental beam width (12.7 $\pm$ 0.4 $\mu$m) and thickness (5.51 $\pm$ 0.03 $\mu$m). Ideally, the sheet thickness should align closely with the depth of field (DOF) of the detection objective to reduce the detection of as much out-of-focus light as possible. Utilizing Eq.~(\ref{eq:DOFobj})
\begin{eqnarray}
DOF = \frac{\lambda n}{NA^2} + \frac{n P_{P}}{M NA},
\label{eq:DOFobj}
\end{eqnarray}
where $\lambda$ is the emission wavelength, $n$ the refractive index, $NA$ the numerical aperture of the objective, $P_{I}$ the image pixel size, and $M$ the objective magnification, for our selected 40X, NA 0.8 water dipping objective, assuming an emission wavelength of 575 nm, we computed a DOF of 1.20 $\mu$m. Given that the DOF is smaller than the beam thickness, some background signal is expected in the images.

The final measured sheet dimensions (12.7 $\pm$ 0.4 $\mu$m by 5.51 $\pm$ 0.03 $\mu$m) closely approach, but slightly exceed, the theoretical values calculated in Sec.~\ref{sec:OurLSM} (11.23 x 4.24 $\mu$m). This deviation likely results from the optics themselves and the physical measurement of the sheet outside the immersion media of the objectives. Given our system is designed for water immersion, measuring the sheet in air compromises resolution and produces a blurrier image on the beam profiling camera.

\section{\label{sec:Data}Data collection and image reconstruction methods for LSM}
When acquiring data at different depths, the end goal of the imaging determines the axial size of the steps between each dataset, known as "z-slices," and the acquisition time. These steps should be sufficiently small for continuous imaging, tailored to the size of the feature of interest. Typically, smaller step sizes (\textit{e.g.}, 10-100 nm) are ideal. However, reconstructing samples with larger features (\textit{e.g.}, 1-1000 $\mu$m) using fine steps could result in excessively large (\textit{e.g.}, 1 TB) data files. Thus, there must be a balance between the number of steps required for accurate reconstruction and managing file size, which varies depending on the imaging method and information sought.

After data collection, the data must be reconstructed and analyzed. For our LSM data, we have developed a 3D image reconstruction code, available on GitHub and accompanied by a user guide.\cite{kisley_research_group_cwru_fcssofilsm_3d_fcssofi_20240425_nodate} Here, we outline the steps for reconstructing spatial information from a 2 $\mu$m bead calibration sample (described in Sec.~\ref{sec:CalStand}) as well as spatiotemporal information collected from 155 kDa dextran diffusing within a 2 wt\% agarose gel.

\subsection{\label{sec:DataWorkup} Data collected on the LSM must be converted and corrected}
The following discussion will begin with the data workup procedure for static samples. In general, the method for static data analysis is simpler than that of the method for dynamic data. Therefore, it provides a good introduction to the overall analysis process.

\subsubsection{\label{sec:Static} Static data requires de-skewing and de-blurring corrections of the raw images}
In our set-up, Micro-Manager is used for data collection, saving raw data files in .tiff format that are subsequently imported into the reconstruction code.\cite{edelstein_computer_2010,edelstein_advanced_2014} Users have the flexibility to import individual z-slices or a complete sequence comprising all collected z-slices. When importing slices individually, they are consolidated into a single sequence and then converted into a unified .mat file format. Before initiating the code, users input specific frames desired and define the region of interest (ROI).

Upon converting the data to .mat format, we address skew, as discussed in Sec.~\ref{sec:CalOffset}, where each subsequent z-slice is offset based on the piezostage position and our calibration data. Our code corrects for skew by shifting each z-slice relative to the location of the first slice in the image sequence. This correction involves translating the image matrix along both x- and y-axes according to the pixel shift measured for a 5 $\mu$m z-step, scaled to account for the actual z-step size between images.

3D deconvolution improves resolution by de-blurring the contribution of the LSM PSF. Users can opt for instant deconvolution or an iterative approach. Instant deconvolution applies a deconvolution algorithm (\textit{e.g.}, Wiener or Richardson-Lucy) once, yielding relatively fast results suitable for real-time applications but may be less accurate with poorly characterized PSFs.\cite{toloui_high_2015,sanders_real_2022}  In contrast, iterative deconvolution assumes the observed image is a convolution of the PSF and real image, starting with an initial guess and adjusting until convergence.\cite{richardson_bayesian-based_1972,zeng_unmatched_2000} In general, iterative approaches are performed within the frequency domain.\cite{kawata_iterative_1980} While iterative methods offer higher accuracy for features that are not point-like, they can be computationally intensive, particularly for large or complex images.\cite{ichioka_iterative_1981,guo_accelerating_2019}

We adapted an iterative code written originally for depth-resolved holographic reconstruction, leveraging its ability to restore continuous, non-bead-like features.\cite{latychevskaia_depth-resolved_2010} The PSF is defined based on the diffraction-limited 2D PSF of the microscope, determined in Sec.~\ref{sec:CalPixel}, and extended to a 3D PSF. Using the MATLAB function 'fspecial3', we generate a 3D PSF based on the 2D PSF and matching the object data file dimensions and z-range. The relationship between the 3D PSF and the unknown 3D object guides the iterative loop, with each subsequent iteration approaching closer towards reconstructing the true 3D object dimensions.

Given the sheet thickness (5.5 $\mu$m) exceeds the DOF (1.44 $\mu$m), some background intensity remains. To enhances image quality, we background correct via thresholding. Our thresholding process, based on sparse emitters, defines an ROI and calculates local background and standard deviation intensities.\cite{shuang_troika_2013} The code iteratively analyzes the entire image and generates a threshold map by adding three standard deviations to the average background intensity.  The corrected frame is generated by subtracting the threshold map from the original frame. This process is performed for each frame in the sequence, ensuring intensity correction for optimal 3D structure reconstruction.

\subsubsection{\label{sec:Dynamic} Dynamic data can be analyzed by correlation to resolve diffusion and structure in 3D}
The dynamic dataset will undergo a similar analysis procedure as the static dataset, with the primary distinction being that instead of containing a single image at each z-step, it now consists of a movie. Prior to de-skewing and deconvolution, the dataset will be transformed into a single image per movie, accomplished through fluorescence correlation spectroscopy super-resolution optical fluctuation imaging (fcsSOFI) analysis.\cite{kisley_characterization_2015,yoshida_computationally-efficient_2021,chatterjee_spatially_2023,antarasen_cross-correlation_2024} fcsSOFI allows for simultaneous quantification of molecule diffusion speeds in heterogeneous media while recovering the matrix structure. Current temporal and structural resolutions of fcsSOFI are approximately 1 $\mu$m$^{2}$s$^{–1}$ and 100 nm, respectively.\cite{chatterjee_spatially_2023}

In our prior publications, we thoroughly explained the theories and methodologies underlying FCS and SOFI correlation analysis techniques. \cite{kisley_characterization_2015,yoshida_computationally-efficient_2021,antarasen_cross-correlation_2024} To summarize, the data of intensity against time at individual pixels undergoes a second-order correlation. For SOFI spatial analysis, cross-correlation is used to reduce aliasing in spatial information.\cite{antarasen_cross-correlation_2024} Subsequently, an image is generated where each new pixel intensity is determined by the cross-correlation function value at the initial time lag. This is unlike conventional optical imaging methods, where reported intensity corresponds to detected photon counts. For FCS diffusion analysis, the correlation curves at each pixel are  fit to pre-selected diffusion model like single- or two-component Brownian or anomalous diffusion.\cite{yoshida_computationally-efficient_2021} Fitting the auto-correlation curves yields the diffusion time $\tau_{D}$ for each pixel, which is then used to calculate the diffusion coefficient $D$ using Eq.~(\ref{eq:Diff})
\begin{eqnarray}
D = \frac{\omega^2}{4\tau_{D}},
\label{eq:Diff}
\end{eqnarray}
where $\omega^2$ is the focal area characterized by the standard deviation of the 2D PSF. A diffusion map is then generated containing $D$ values for each pixel. The fcsSOFI image is then formed in an image fusion step of the SOFI image indicating the saturation and the FCS image indicating the hue for a final super-resolved image of diffusion rates at each 2D image size.

\subsection{\label{sec:3DRecon} Intensity corrected z-slices can be stacked to recreate the 3D structure}
\begin{figure}
\includegraphics[width=0.5\textwidth]{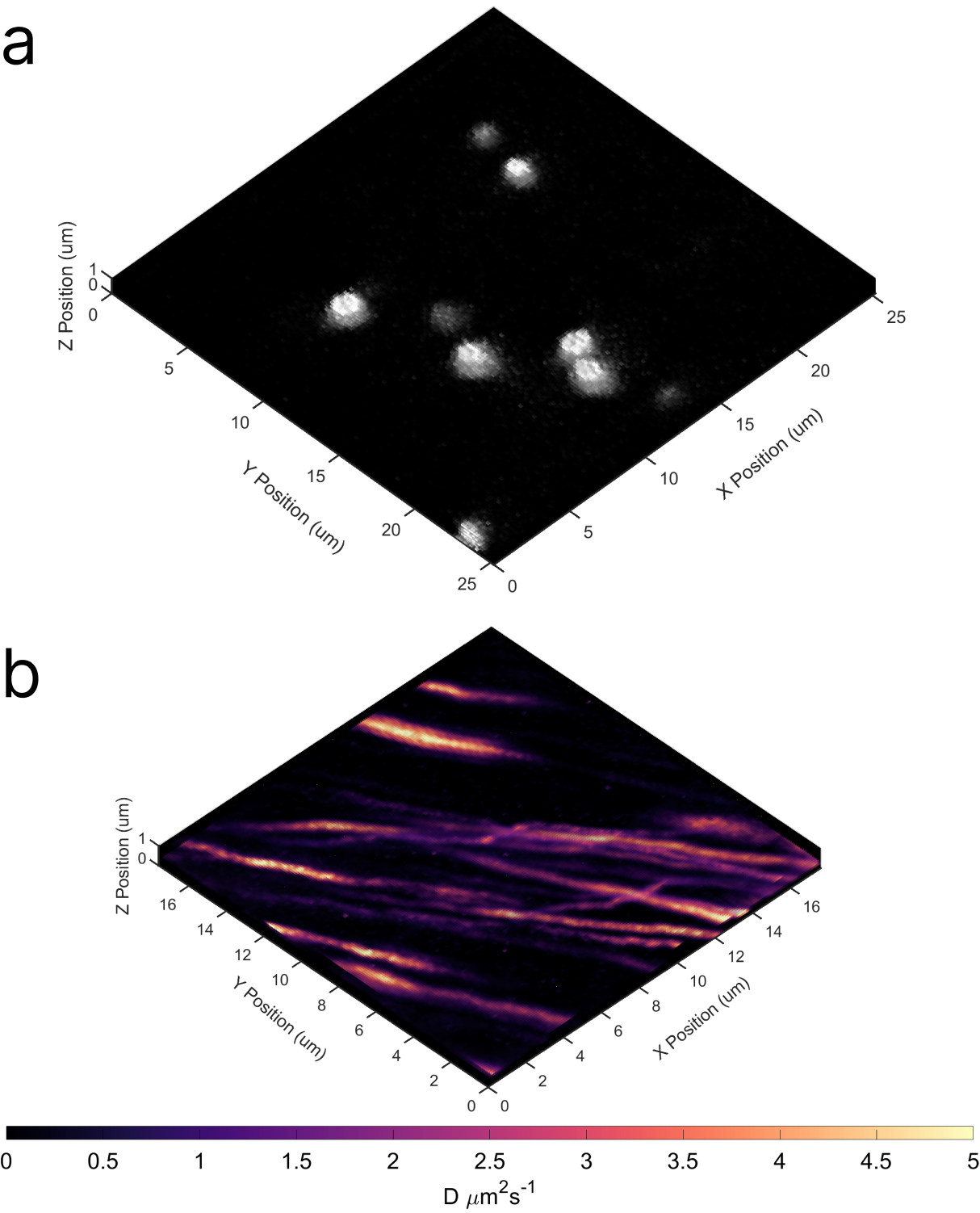}
\caption{\label{fig:3D}3D reconstruction of stationary 2 $\mu$m beads (a) and dynamic 155 kDa dextran (b) in a 2 wt\% agarose gel generated using our 3D reconstruction code from data collected on our home-built LSM.}
\end{figure}

The final corrected images are stacked to reconstruct the 3D volumetric image. This process is accomplished in our code using the function 'vol3d'.\cite{woodford_vol3d_2024} Briefly, the 'vol3d' function creates a volume render from the inputted 3D z-slice sequence. The resulting 3D image (Fig.~\ref{fig:3D}) is scaled in the xy-plane based on a factor of the number of pixels the intensities cover and the value of $P_{I}$. The z-axis scaling is based on a factor of the number of slices and the physical step between them.

Starting with the reconstruction of the static 2 $\mu$m bead data (Fig.~\ref{fig:3D}a), we see a heterogeneous distribution of the beads within the hydrogel environment. Performing line section analysis on the beads in view (Fig.~S5) reveals the average diameter is 1.6 $\pm$ 0.5 $\mu$m (x-axis) and 1.7 $\pm$ 0.7 $\mu$m (y-axis), slightly less than the manufacturer reported size (2.10 $\pm$ 0.09 $\mu$m, Table~\ref{tab:PSFbeads}), but still within variation. The lower average diameter is most likely caused by the deconvolution step, which is known to over-correct the image. Additionally, there is variability in the bead locations within the light-sheet with beads ranging from partially sectioned to fully illuminated by the sheet. This effect is demonstrated in Fig.~\ref{fig:3D}a through the variability in saturation where beads that are partially in the light-sheet are less intense than those that are fully within.

Next, we examine the validity of using our home-built LSM to measure spatiotemporal information from 155 kDa dextran diffusing within a 2 wt\% agarose gel. Here, we rely on the complex porous structure of agarose  for our nanoscale imaging. These hydrogels are known to have a wide range of structural features including isolated cavities and interconnected channels with an average size of approximately 150 nm.\cite{pluen_diffusion_1999,khodadadi_yazdi_agarose-based_2020}
When allowed to freely diffuse in water, the diffusion coefficient of 155 kDa dextran should be around 24 $\mu$m$^{2}$s$^{-1}$ given the molecule has a hydrodynamic radius of $\sim$9 nm.\cite{kalwarczyk_apparent_2017} Introducing a porous structure should induce a confinement effect, reducing the overall diffusion coefficient of the molecule. Previously, we have shown 2 wt\% agarose slows the diffusion of 155 kDa dextran to 2.8 $\pm$ 0.4 $\mu$m$^{2}$s$^{-1}$. However, we noted this speed is slower than expected and suspect interactions at the glass interface to be the cause.\cite{chatterjee_spatially_2023} Here, the average diffusion coefficient of our 3D system is 4.8 $\pm$ 0.7 $\mu$m$^{2}$s$^{-1}$ (Fig.~\ref{fig:3D}b), faster than that of our previously reported TIRF results due to the lack of interactions at an interface. We have also previously reported the pore structure of the agarose gel, albeit limited to the glass interface, using our fcsSOFI technique.\cite{chatterjee_spatially_2023} Here, we are able to recover not only isolated cavities, but also interconnected channels (Fig.~\ref{fig:3D}b). Performing line section analysis, we find the average cross-section is 340 $\pm$ 50 nm (Fig.~S6).

\section{\label{sec:Concl}Conclusion}
This paper provides an accessible practical guide to LSM for scientists new to this technique. We emphasize that LSM is suitable not only for imaging static microscale features but also nanoscale features and diffusion dynamics. We cover fundamental aspects of microscope conception, beam profiling, and guide readers through the entire process, concluding with image reconstruction.
In Sec.~\ref{sec:Config}, we discuss essential practical considerations for designing a home-built LSM system. This includes decisions on beam profiling, static or dynamic sheet formation, objectives, sample mounting, and camera selection. We highlight the importance of objective orientation, as this choice impacts the overall design and adaptability of LSM for various samples and imaging conditions.
Following component selection, Sec.~\ref{sec:Alignment} outlines our home-built set-up to illustrate the alignment and calibration process. While alignment procedures are similar to other optical microscopes, LSM calibration differs due to its 3D imaging capability and axial skewing during sample movement within the light-sheet. Correcting this offset is crucial for accurate 3D image reconstruction. Given the custom nature of our system, we detail the conditions and steps required for constructing a continuous 3D image using our home-built software in Sec.~\ref{sec:Data}.
Each LSM system is unique, presenting challenges in design and construction tailored to specific research needs. We hope this guide serves as a valuable resource, offering clarity and guidance in the complex field of light-sheet microscopy.

\section*{Supplementary Material}
Methods including sample preparations and imaging conditions, skew correction calculations, photograph of our LSM set-up (S1), photograph of a highly fluorescent bulk bead standard solution (S2), time-lapse line sections of 2 $\mu$m bead (S3), image of 2 $\mu$m bead to demonstrate the tilt of the light-sheet (S4), line sections of a 2 $\mu$m bead after 3D
reconstruction (S5), cross-section of an agarose pore (S6), graphical representation of the light-sheet geometry (S7), all components used for our LSM set-up (Table S1), and legends for CAD Files S1 to S7.(PDF) 

CAD S1: Bath Drawing (PDF)

CAD S2: Microscope Sample Holder Assembly (PDF)

CAD S3: Microscope Stage Mount Knob (STL)

CAD S4: Microscope Stage Mount Flat Pattern of Slide Holder (DXF)

CAD S5: Microscope Stage Mount Riser (STL)

CAD S6: Microscope Stage Mount Base (DXF)

CAD S7: Laboratory Jack to Stepper Stage Adapter (PDF)

\begin{acknowledgments}
We acknowledge NIH NIGMS grant R35GM142466 and the Paul G. Allen Family Foundation's Allen Distinguished Investigators Award for financial support of this work. The authors also thank the Kisley research group, the Jenkins research group and Matthew Kramer for the helpful discussions.
\end{acknowledgments}

\section*{Author Declarations}
%\subsection*{Conflict of Interest}
The authors have no conflicts to disclose.

\section*{Data Availability Statement}
The data that support the findings of this study are available within the supplementary material and also available upon reasonable request.

%\section*{References}
\nocite{*}
\bibliography{main}

\end{document}